\documentclass{emulateapj}

\begin{document}

\title{Probing the Relation Between X-ray-Derived and Weak-Lensing-Derived Masses for Shear-Selected Galaxy Clusters: I. A781}

\author{Neelima Sehgal, John P. Hughes}
\affil{Department of Physics and Astronomy, Rutgers University, 136
Frelinghuysen Road, Piscataway, NJ 08854-8019}
\author{David Wittman, Vera Margoniner, J. Anthony Tyson, Perry Gee}
\affil{Department of Physics, University of California, Davis, 1
Shields Avenue, Davis, CA 95616}
\author{Ian dell'Antonio}
\affil{Department of Physics, Brown University, Providence, RI
02912}

\begin{abstract}

We compare X-ray and weak-lensing masses for four galaxy clusters
that comprise the top-ranked shear-selected cluster system in the
Deep Lens Survey.  The weak-lensing observations of this system,
which is associated with A781, are from the Kitt Peak Mayall 4-m
telescope, and the X-ray observations are from both \emph{Chandra}
and \emph{XMM-Newton}. For a faithful comparison of masses, we adopt
the same matter density profile for each method, which we choose to
be an NFW profile. Since neither the X-ray nor weak-lensing data are
deep enough to well constrain both the NFW scale radius and central
density, we estimate the scale radius using a fitting function for
the concentration derived from cosmological hydrodynamic simulations
and an X-ray estimate of the mass assuming isothermality.  We keep
this scale radius in common for both X-ray and weak-lensing
profiles, and fit for the central density, which scales linearly
with mass.  We find that for three of these clusters, there is
agreement between X-ray and weak-lensing NFW central densities, and
thus masses.  For the other cluster, the X-ray central density is
higher than that from weak-lensing by 2$\sigma$. X-ray images
suggest that this cluster may be undergoing a merger with a smaller
cluster.  This work serves as an additional step towards
understanding the possible biases in X-ray and weak-lensing cluster
mass estimation methods. Such understanding is vital to efforts to
constrain cosmology using X-ray or weak-lensing cluster surveys to
trace the growth of structure over cosmic time.

\end{abstract}


\section{Introduction}

Galaxy clusters have the potential to open a new window on cosmology
by serving as precision tracers of the growth of structure over
cosmic time. The growth of structure can provide independent
constraints on the matter density ($\Omega_{\textrm{\tiny{M}}}$),
the dark energy density ($\Omega_{\Lambda}$), and the dark energy
equation of state ($\omega$), that would both verify our standard
cosmological picture and take us further into understanding the
nature of dark energy (e.g., \citet{Carlstrom2002}). Utilizing
galaxy clusters as tracers of structure growth largely relies on
knowledge of cluster masses. Ideally cluster samples would have
selection criteria based on mass, and mass estimates of clusters
would be based on probes of their gravitational potential. However,
most large samples of clusters that exist to date are selected on
the basis of their trace baryons (i.e., visible light from galaxies
or X-ray emission from hot intracluster gas). Moreover, traditional
probes of cluster mass (X-ray and optical) depend on the cluster's
star formation history, baryon content, and assumptions about its
dynamical state. Only recently have we obtained samples of clusters
of significant size unbiased with respect to baryons and instead
selected on the basis of their weak gravitational lensing shear.

One such sample is provided by the Deep Lens Survey (DLS), a deep
BVR\emph{z}$^\prime$ imaging survey of 20 square degrees
\citep{Wittman2002}.  The observations were taken with the Cerro
Tololo Blanco and Kitt Peak Mayall 4-m telescopes. The primary goal
of this survey is to study the growth of mass clustering over cosmic
time using weak lensing. The DLS team has shown it is capable of
finding new galaxy clusters using their weak-lensing signal alone
\citep{Wittman2001, Wittman2003}, and it has presented its first
sample of cluster candidates from the first 8.6 square degrees of
the survey \citep{Wittman2006}. The DLS survey should find $\sim$40
clusters when completed.  The CFHT Legacy Survey Deep has also
presented shear-selected clusters from a 4 square degree region and
the Garching-Bonn Deep Survey has presented a sample from 19 square
degrees \citep{Gavazzi2007, Schirmer2007}.

We have been following-up a shear-ranked sample of DLS clusters with
\emph{Chandra} and \emph{XMM-Newton}. One goal of this X-ray
follow-up is to confirm that the DLS shear-selected cluster
candidates are in fact true virialized collapsed structures.
Preliminary analysis for five of these clusters is presented in
\citet{Hughes2004}. A further goal of this X-ray follow-up is to
characterize the robustness of X-ray and weak-lensing cluster mass
estimates and the biases inherent in X-ray and shear-selected
samples. This understanding is necessary in order to lay the
groundwork for precision cosmology via larger X-ray and weak-lensing
cluster surveys (utilizing, for example, Constellation-X
\footnote{http://constellation.gsfc.nasa.gov/} and LSST
\footnote{http://www.lsst.org/lsst\_home.shtml}). Such
characterizations are facilitated by comparing weak-lensing mass
estimates with X-ray mass estimates, as we elaborate on in \S2.

Below we report our weak-lensing and X-ray mass estimates for our
top ranked shear-selected cluster, A781, and three surrounding
clusters. We discuss details of the weak-lensing and X-ray
observations in \S3, and the details of the weak-lensing and X-ray
mass estimation methods in \S4 and \S5. In \S6 we discuss our
results, and in \S7 we summarize our conclusions.

\section{Benefits of Investigating the Relation Between X-ray- and Weak-Lensing-Derived Masses for Shear-Selected Clusters}

An important issue for shear-selected clusters is projection bias.
The weak-lensing shear signal is sensitive to all the intervening
matter between the background galaxies and the observer. This leads
to a possible projection bias of shear mass estimates as non-cluster
line-of-sight matter contaminates the shear signal (e.g.,
\citet{Metzler1999, White2002, dePutter2005}).  X-ray observations
provide an independent way to estimate the mass.  Thus to quantify
the extent of this contamination, we wish to compare X-ray mass
estimates to weak-lensing mass estimates for clusters that are
dynamically relaxed. X-ray observations are uniquely suited to this
because they offer clear indications of a cluster's dynamical state
via X-ray images and temperature measurements. We impose the
condition of relaxation because X-ray mass estimates are based on an
assumption of hydrostatic equilibrium and are likely invalid for
highly unrelaxed systems. The comparison of X-ray to weak-lensing
mass estimates will indicate how significantly projection bias
affects the latter.

To study the bias (or absence of bias) inherent in X-ray and
shear-selected cluster surveys, we first note that optical selection
depends on star formation history and X-ray/Sunyaev-Zel'dovich
selection depends on the heating of the intracluster medium.  It has
been proposed that up to 20$\%$ of shear-selected clusters have not
yet heated their intracluster medium enough to be visible by current
X-ray satellites \citep{Weinberg2002}.  This is because a
significant fraction of cluster-mass overdensities are likely
nonvirialized and still in the process of gravitational collapse.
These nonvirialized overdensities should produce much weaker X-ray
emission than that from a fully virialized cluster of the same mass.
Differentiating this population from false-positive shear signals
due to unrelated line-of-sight projections that appear as single
larger mass concentrations, will be a challenge.  Such a `dark lens'
cluster candidate was reportedly found by \citet{Erben2000} via
weak-lensing observations centered on Abell 1942.  This detection
was followed-up by \citet{Gray2001} in the infrared with no obvious
luminous counterpart detected.  Several more apparent `dark lenses'
are reported in \citet{Koopmans2000}, \citet{Umetsu2000}, and
\citet{Miralles2002}. If such `dark lenses' exist, there should
exist a continuum of clusters between those which just satisfy
M$_{\textrm{xray}}$ $<$ M$_{\textrm{weaklens}}$ and those which
simply show no detectable X-ray counterpart to their weak-lensing
signal. Characterizing and quantifying the clusters for which
M$_{\textrm{xray}}$ $<$ M$_{\textrm{weaklens}}$ will allow greater
understanding of which clusters are missed by traditional samples
and the percentage of false-positive detections that are inherent in
shear surveys.

The ratio of M$_{\textrm{xray}}$/M$_{\textrm{weaklens}}$ may also
prove to be a good diagnostic of the dynamical relaxation of a
cluster.  Recent findings based on 22 high X-ray luminosity,
low-redshift (0.05 $<$ z $<$ 0.31) clusters, selected on the basis
of their high X-ray emission and targeted for weak-lensing follow-up
with the ESO VLT, suggest X-ray cluster mass estimates larger than
weak-lensing mass estimates positively correlate with clusters being
dynamically unrelaxed \citep{Cypriano2004}. Naively one would expect
this theoretically because events (such as mergers) that disrupt a
cluster's equilibrium introduce transient shock heating of its
intracluster gas. Calculating X-ray cluster masses using an
assumption of hydrostatic equilibrium and a higher temperature than
the cluster would have if relaxed, results in an overestimate of the
true mass. However, recent work based on hydrodynamic cluster
simulations suggests X-ray mass estimates are biased low for
unrelaxed clusters because only a portion of the kinetic energy of
the merging system is converted into thermal energy of the
intracluster medium, for even an advanced merger, while the mass of
the merging system has already increased
\citep[e.g.,][]{Kravtsov2006}.  Comparing M$_{\textrm{xray}}$ to
M$_{\textrm{weaklens}}$ for our shear-selected clusters would
determine whether X-ray mass estimates are biased high or low for
unrelaxed clusters and whether this ratio can be used as a universal
diagnostic of cluster dynamical state. Such a universal diagnostic
would prove useful in investigating cluster evolution.

Finally, there have been several reported instances of clusters that
have an X-ray signal but no apparent weak-lensing counterpart
\citep{Cypriano2004, Dahle2002}.  We have detected such a cluster
while following-up our highest shear-ranked cluster with
\emph{XMM-Newton}. This cluster did not appear in the original shear
maps made for the DLS survey but is readily apparent in
\emph{XMM-Newton} observations. The inverse of `dark lenses',
negative weak-lensing detections are not unexpected since weak
lensing is a less sensitive method of cluster searching as many
galaxies need to be detected behind a cluster.  Also mergers could
potentially boost the X-ray signal of clusters otherwise below both
current X-ray and weak-lensing thresholds.  It is important for
understanding the limitations of weak-lensing surveys to explore
what is occurring in cases such as these.

\section{Observations}

\subsection{Weak-Lensing Observations}

The Deep Lens Survey consists of five fields, each $2^\circ\times
2^\circ$ and isolated from each other.  The two northern fields were
observed using the Kitt Peak Mayall 4-m telescope, and the three
southern fields were obtained with the Cerro Tololo Blanco 4-m
telescope. Observing began in November 1999 at Kitt Peak and in
March 2000 at Cerro Tololo.  The deep BVR\emph{z}$^\prime$ images
were taken with 8k $\times$ 8k Mosaic imagers \citep{Muller1998} on
each telescope, which provided 35$^\prime\times 35^\prime$ fields of
view with 0.26$^{\prime\prime}$ pixels and minimal gaps between the
CCD devices. The observing strategy was to require better than
0.9$^{\prime\prime}$ seeing in the R band, so that this band would
have good, largely uniform resolution. When the seeing was worse
than this, B, V, and \emph{z}$^\prime$ images were taken.  The
source galaxy shapes were measured in the R band, and B, V, and
\emph{z}$^\prime$ images provided color information and photometric
redshifts. \citet{Wittman2002} gives details of the field selection
and survey design, and \citet{Wittman2006} gives details regarding
the image processing and convergence maps.

A list of cluster candidates was compiled, based on the first 8.6
deg$^{2}$ of processed DLS data, and the candidates were ranked by
their shear peak values.  Multiple peaks within a 16$^\prime$ box
were considered a single target for purposes of \emph{Chandra}
follow-up. A781 emerged as the top-ranked cluster candidate, with
both DLS and archived \emph{Chandra} observations indicating that
this cluster was really a complex of several clusters
\citep{Wittman2006}.  X-ray and optical follow-up of the A781
cluster complex was pursued as part of a larger follow-up program
that will encompass a significant sample of DLS cluster candidates.

\subsection{X-ray Observations}

We were awarded 15ks of \emph{XMM-Newton} time in cycle 2 to get a
closer look at our top-ranked DLS cluster complex.  This observation
took place on 04 April 2003 (Obsid$\#$ 0150620201).  In addition,
\emph{Chandra} had observed A781 on 03 October 2000 with the ACIS-I
detector for a nominal exposure time of 10 ks (Obsid \# 534). The
\emph{XMM-Newton} and \emph{Chandra} observations revealed that the
A781 cluster complex consists of a large main cluster connected to a
subcluster with two smaller clusters to its east and one to its
west.  We shall call the largest cluster the `Main' cluster.  The
subcluster to its southwest appears in the act of merging with it
(see Figure \ref{fig:images}).  Just to the east of the Main cluster
is another cluster, which we will refer to as the `Middle' cluster,
and within the same pointing, further to the east, is another
cluster, hereafter `East' cluster. The \emph{XMM-Newton} observation
also presented us with a surprise. To the west of the Main cluster
there appears to be one more cluster, which we will call the `West'
cluster.  This cluster did not appear in the original DLS
convergence maps made for the survey, and it is also, unfortunately,
out of the field of view of the \emph{Chandra} archive observations.
Table \ref{tab:IAUnames} lists the IAU designations of these
clusters.

\begin{table}
\begin{center}
\begin{tabular}{|c|c|}
\hline  IAU Designation&Nickname\\
\hline \hline
\hline CXOU J092026+302938&Main Cluster\\
\hline CXOU J092053+302800&Middle Cluster\\
\hline CXOU J092110+302751&East Cluster\\
\hline CXOU J092011+302954&Subcluster\\
\hline XMMU J091935+303155&West Cluster\\
\hline
\end{tabular}
\end{center}
\caption{IAU Designations for the clusters in the A781 cluster
complex.} \label{tab:IAUnames}
\end{table}

\subsection{Optical Spectroscopy}

\citet{Geller2005} conducted a magnitude-limited (to $R=20.5$)
spectroscopic survey in this field.  They report mean redshifts of
0.302, 0.291, and 0.427 for the Main, Middle, and East clusters
respectively (labeled as clusters A, B, and C in
\citet{Geller2005}). These redshifts were obtained from 163, 123,
and 33 cluster members respectively.  No redshift errors are quoted,
however given the number of cluster members and the redshift errors
in the individual galaxies, the systemic redshifts of the systems
should be accurate to dz of 0.0002. The rest frame line-of-sight
velocity dispersions were found to be $\sigma_A =674^{+43}_{-52}$ km
s$^{-1}$ (Main), $\sigma_B =741^{+35}_{-40}$ km s$^{-1}$ (Middle),
and $\sigma_C =733^{+77}_{-112}$ km s$^{-1}$ (East)
\citep{Geller2005}. According to these velocity dispersions, these
cluster components appear to be similar, a result we examine further
using our X-ray and weak-lensing data.

The redshift of the West cluster is not reported in
\citet{Geller2005}.  We obtained spectroscopy of this cluster with
Keck/LRIS \citep{Oke1995} in longslit mode on 16 January 2007. We
obtained secure redshifts for two member galaxies, at a mean
redshift of $0.428\pm0.001$, though clearly the quoted error is
itself highly uncertain with only two members.  We also observed the
East cluster in longslit mode on the same night, finding a mean
redshift of $0.426\pm0.003$ based on two members, in agreement with
\citet{Geller2005}.  Thus the East and West clusters are the only
two components at the same redshift, but with a transverse
separation of 21$^\prime$ or 7.0 Mpc.  Throughout this work we
assume a $\Lambda$CDM cosmology of $h=0.71$,
$\Omega_{\Lambda}=0.73$, and $\Omega_M=0.27$ \citep{Spergel2003}.

\section{Extracted X-ray Temperatures and Gas Density
Profiles}\label{sec:profiles}

To obtain X-ray mass estimates of these clusters, we follow the
standard practice of treating the intracluster gas as a hydrostatic
fluid.  This assumption is reasonable for dynamically relaxed
clusters since collision times for ions and electrons in the hot gas
are very short compared to times scales over which the gas heats or
cools or the cluster gravitational potential varies.  Assuming
spherical symmetry,
\begin{eqnarray}
M(r)=-\frac{krT(r)}{G\mu m_p}\Big(\frac{d\ln\rho(r)}{d\ln
r}+\frac{d\ln T(r)}{d\ln r}\Big), \label{eq:hydrostatic}
\end{eqnarray}
where $\mu m_p$ is the gas mean molecular weight, $T(r)$ and
$\rho(r)$ are the gas temperature and density profiles, and $M(r)$
is the total mass within a radius $r$.  We assume the cluster gas
follows a $\beta$-model,
\begin{eqnarray}
\rho(r) =
\rho_{0_g}\big[1+\big(\frac{r}{r_c}\big)^2\big]^{-\frac{3\beta}{2}},
\label{eq:beta_model}
\end{eqnarray}
proposed by \citet{Cavaliere1978}.  In this model, the core radius,
$r_c$, is where the density is half the central density for a
typical $\beta$ of 2/3.  To determine $\beta$ and $r_c$, we note
that the X-ray emissivity is proportional to the square of the
cluster gas density times the cooling function, i.e., $\epsilon (r)
\propto \Lambda (T(r)) (n_e (r))^2$. Integrating the emissivity
through the cluster line of sight gives the X-ray surface brightness
$\Sigma$. Assuming a $\beta$-model for the gas density and noting
that the cooling function is close to constant over the range of
typical cluster temperatures yields
\begin{eqnarray}
\Sigma (b) \hspace{0.15cm}= \hspace{0.15cm}
\int_{-\infty}^{\infty}\epsilon (r)dl \hspace{0.15cm} \propto
\hspace{0.15cm}
\bigg(1+\Big(\frac{b}{r_c}\Big)^2\bigg)^{-3\beta+\frac{1}{2}},
\label{eq:surface_brightness}
\end{eqnarray}
where $l=\sqrt{r^2-b^2}$ and $b$ is the projected radius (e.g.,
\citet{Sarazin1986}).  We fit the radial X-ray surface brightness
profile with equation \ref{eq:surface_brightness} to obtain
estimates for $\beta$ and $r_c$ (see \S4.2).

The statistical quality of our data precludes the determination of a
radial temperature profile.  In lieu of this, we assume an NFW
profile for the cluster matter density given by
\begin{eqnarray}
\rho_M (r) = \frac{\rho_0}{(r/r_s)(1+r/r_s)^2} \label{eq:NFW}
\end{eqnarray}
where $\rho_0$ is the central density and $r_s$ is the scale radius
\citep{Navarro1996,Navarro1997}.  This gives
\begin{eqnarray}
M(r)=4\pi \rho_0 r_s^3 \Big(\textrm{ln}(1+\frac{r}{r_s}) +
\frac{1}{1+r/r_s} - 1\Big) \label{eq:NFW_mass}
\end{eqnarray}
for the mass within a radius r.  Using equation
\ref{eq:hydrostatic}, we solve for the cluster temperature profile
and then for the emission-weighted, projected, average temperature
within a given aperture, which our data allow us to measure
relatively precisely.  We compare this predicted temperature to our
measured temperature determined from X-ray spectroscopy to find the
best-fit value for $\rho_0$ and thus the cluster mass.  We describe
the details of the X-ray spectroscopy below.

\begin{table*}
\begin{center}
\begin{tabular}{|c|c|c|}
\hline  \emph{Chandra}&source&background annulus (same center as source)\\
\hline \hline
\hline Main Cluster&09:20:24.8 +30:30:20.4  2.4'&3.3' -- 4.3' (minus Subcluster)\\
\hline Middle Cluster&09:20:52.5 +30:28:08.4  1.6'&2.3' -- 3.3' (minus East)\\
\hline East Cluster&09:21:10.9 +30:28:04.2  1.5'&2.0' -- 3.0' (minus Middle)\\
\hline Subcluster&09:20:09.4 +30:30:02.5 0.9'&...\\
\hline
\end{tabular}
\end{center}
\caption{\emph{Chandra} cluster and background annulus extraction
regions.  Point sources excluded from the cluster regions are given
in Table \ref{tab:Chandra_points}.}
\label{tab:Chandra_cluster_regions}
\end{table*}

\begin{table*}
\begin{center}
\begin{tabular}{ccc}
\hline R.A. & Decl. & Radius \\
\hline\hline
09:20:32.590 & +30:29:10.49 & 4" \\
09:20:29.774 & +30:28:55.73 & 2" \\
09:20:59.852 & +30:27:29.23 & 8" \\
09:20:50.765 & +30:29:22.97 & 5" \\
09:20:45.587 & +30:28:39.22 & 8" \\
09:21:06.817 & +30:27:40.20 & 11" \\
09:21:05.684 & +30:29:02.13 & 6" \\
\hline
\end{tabular}
\end{center}
\caption{Point sources excluded from the \emph{Chandra} cluster
regions.} \label{tab:Chandra_points}
\end{table*}

\begin{table*}
\begin{center}
\begin{tabular}{|c|c|}
\hline  \emph{XMM-Newton}&region\\
\hline \hline
\hline Main Cluster&09:20:24.439  +30:30:21.12  2.5'\\
\hline Middle Cluster&09:20:52.433  +30:28:12.74  1.4'\\
\hline East Cluster&09:21:09.912  +30:27:58.31  1.1'\\
\hline Subcluster&09:20:10.046  +30:29:57.17  1'\\
\hline West&09:19:34.752  +30:32:00.88  1'\\
\hline Background Annulus&09:20:12.521  +30:29:10.37  10.7' -- 13.7'
(minus East)\\
 \hline
\end{tabular}
\end{center}
\caption{\emph{XMM-Newton} cluster and background annulus extraction
regions. Point sources excluded from the cluster regions are given
in Table \ref{tab:XMM_points}.} \label{tab:XMM_cluster_regions}
\end{table*}

\begin{table*}
\begin{center}
\begin{tabular}{ccc}
\hline R.A. & Decl. & Radius \\
\hline\hline
09:20:33.209 & +30:28:58.48 & 17" \\
09:20:24.600 & +30:33:19.12 & 24" \\
09:20:21.809 & +30:30:35.14 & 16" \\
09:20:34.805 & +30:29:47.02 & 12" \\
09:20:25.531 & +30:33:39.11 & 12" \\
09:20:45.588 & +30:28:39.22 & 12" \\
09:21:06.816 & +30:27:40.20 & 11" \\
\hline
\end{tabular}
\end{center}
\caption{Point sources excluded from the \emph{XMM-Newton} cluster
regions.} \label{tab:XMM_points}
\end{table*}

\subsection{X-ray Temperatures}

\subsubsection{Analysis of Chandra Data}

\begin{figure*}
\begin{center}
\hbox{
\includegraphics[angle=0,width=8.5cm]{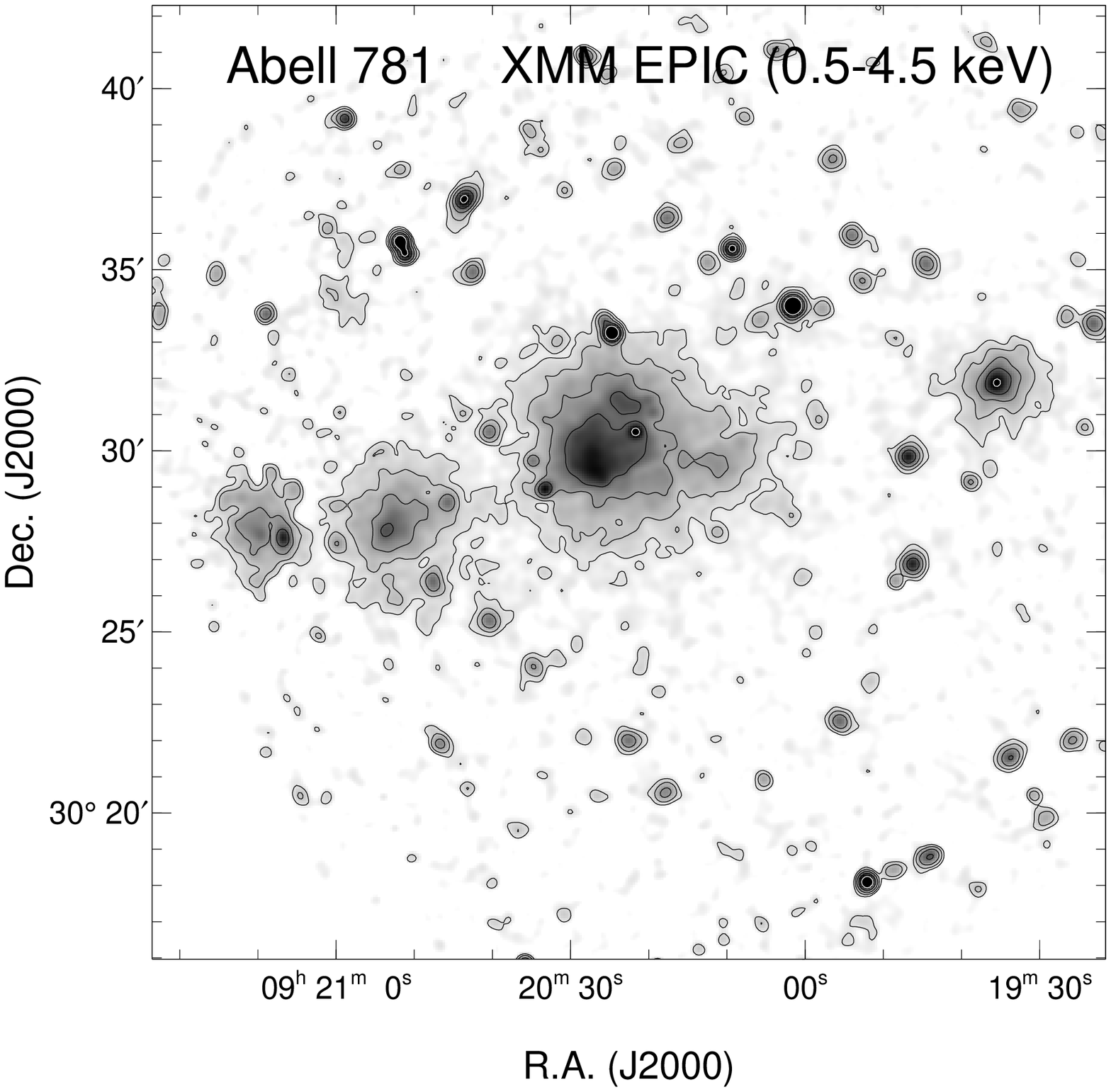}
\includegraphics[angle=0,width=8.5cm]{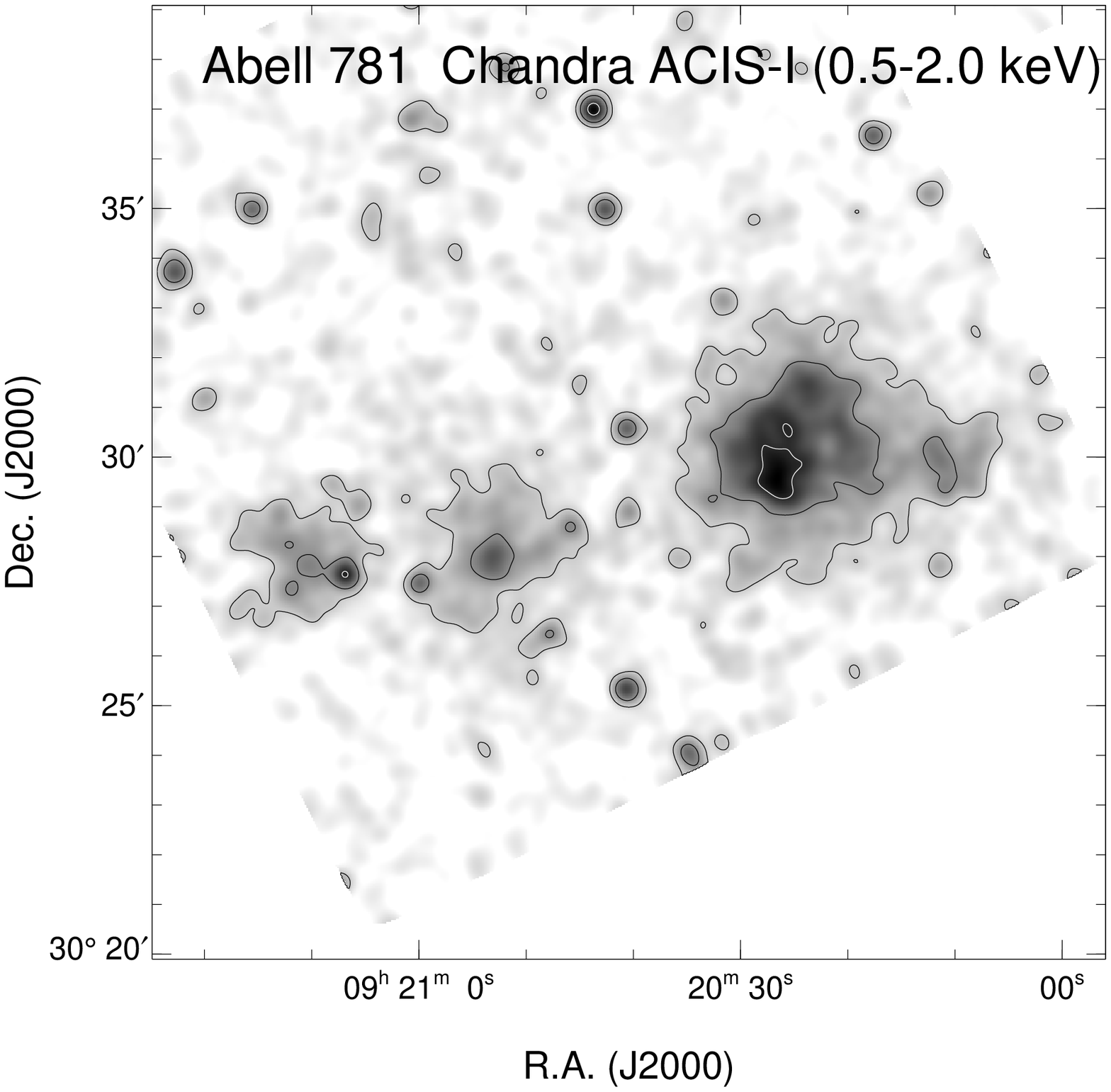}}
\caption{\emph{Left panel}: \emph{XMM-Newton} 15 ks image of the
A781 cluster complex. We refer to these clusters from left to right
as East, Middle, Main, and West. The contours represent $1\times
10^{-3}$ (white), $4.5\times 10^{-4}$, $2\times 10^{-4}$, $8.9\times
10^{-5}$, and $4\times 10^{-5}$ cts/s/(4" square pixel). \emph{Right
panel}: \emph{Chandra} 10 ks image of the A781 cluster complex.  The
contours represent $2.7\times 10^{-5}$ (white), $9.7\times 10^{-6}$,
and $3.5\times 10^{-6}$ cts/s/(2" square pixel). Note the smaller
field of view of the \emph{Chandra} image, which only covers the
three clusters to the east.} \label{fig:images}
\end{center}
\end{figure*}

We downloaded the \emph{Chandra} data from the archive and used CIAO
software tools for the initial data reduction steps. The observation
was carried out in full-frame timed exposure mode using very faint
telemetry mode. The peak X-ray emission of the Main cluster was
positioned near the center of chip I3, some 3.5$^\prime$ from the
on-axis ``sweet'' spot of the high resolution mirror.  The merging
subcluster to the west was imaged on chip I3, while the two other
clusters toward the east were imaged on chip I1. There was no
cluster X-ray emission visible on the remaining two chips of the
imaging array. We note that chip S2 of the spectroscopic array was
also active during this observation, but we did not utilize these
data.  Event pulse heights were corrected for time-dependent gain,
and all grades, other than 0, 2, 3, 4, and 6, were rejected.
Information contained in the very faint mode data was used to reject
non--X-ray background events. The light curve of the entire imaging
array (minus obvious cluster and unresolved emission) was examined
and no time intervals of high or excessive background were found.
The resulting live-time corrected exposure time was roughly 9900
sec.  Figure \ref{fig:images} shows the 500-2000 eV band image of
the ACIS-I data after exposure correction (for which the vignetting
function was calculated at a monochromatic X-ray energy of 1 keV)
and smoothing with a Gaussian of $\sigma \approx 10^{\prime\prime}$.

Spectral extraction regions for the clusters were determined (see
Tables \ref{tab:Chandra_cluster_regions} and
\ref{tab:Chandra_points}) to optimize the signal to noise ratio of
the resulting spectra. Annular regions surrounding each cluster were
used to generate background spectra. Obvious point sources were
excluded from both source and background regions, and cluster
emission was excluded from all background regions. Weighted spectral
response functions were generated for each source and matching
background region, including instrumental absorption due to
contamination build-up on the ACIS filters.

Xspec (version 11.3) was used for the spectral analysis. Fits were
first done to the background spectra using a phenomenological model
consisting of a non-X-ray background component (three gaussian lines
and a power law to account for instrumental fluorescence lines and
charged particles) and an astrophysical component (an absorbed
power-law model to account for the unresolved X-ray background).
Inclusion of a soft thermal component (from nearby diffuse Galactic
emission, for example) did not significantly improve the background
fits, so it was not included.  The absorption column density was
fixed to a value of $N_H = 1.94\times 10^{20}$ atoms cm$^{-2}$ based
on Galactic HI measurements in this direction \citep{Dickey1990} and
the photon index of the astrophysical background component was fixed
to $\Gamma = 1.4$. Only the normalization of this power-law model
was allowed to vary.  For the non-X-ray background, the gaussian
line centroids and normalizations as well as the power-law index and
normalization were allowed to vary freely.  There were a total of
nine free parameters for the fits to the background spectra.  The
source spectra included a redshifted thermal plasma model (mekal, in
xspec parlance) to account for the cluster emission as well as the
full component of background models just described. In all cases
best fits were determined using the ``c-stat'' fit statistic, which
is a likelihood figure of merit function appropriate for
Poisson-distributed data.

Each pair of matched source and background spectra was fitted
jointly with the background spectral components scaled between the
source and background based on the ratio of exposure integrated over
the extraction regions.  In the joint fits the normalizations of the
two background power laws plus the non-X-ray background photon index
were allowed to vary. The other background model parameter values
were held fixed at values determined from the background fits alone.

\begin{figure}
\begin{center}
\includegraphics[angle=0,width=8cm]{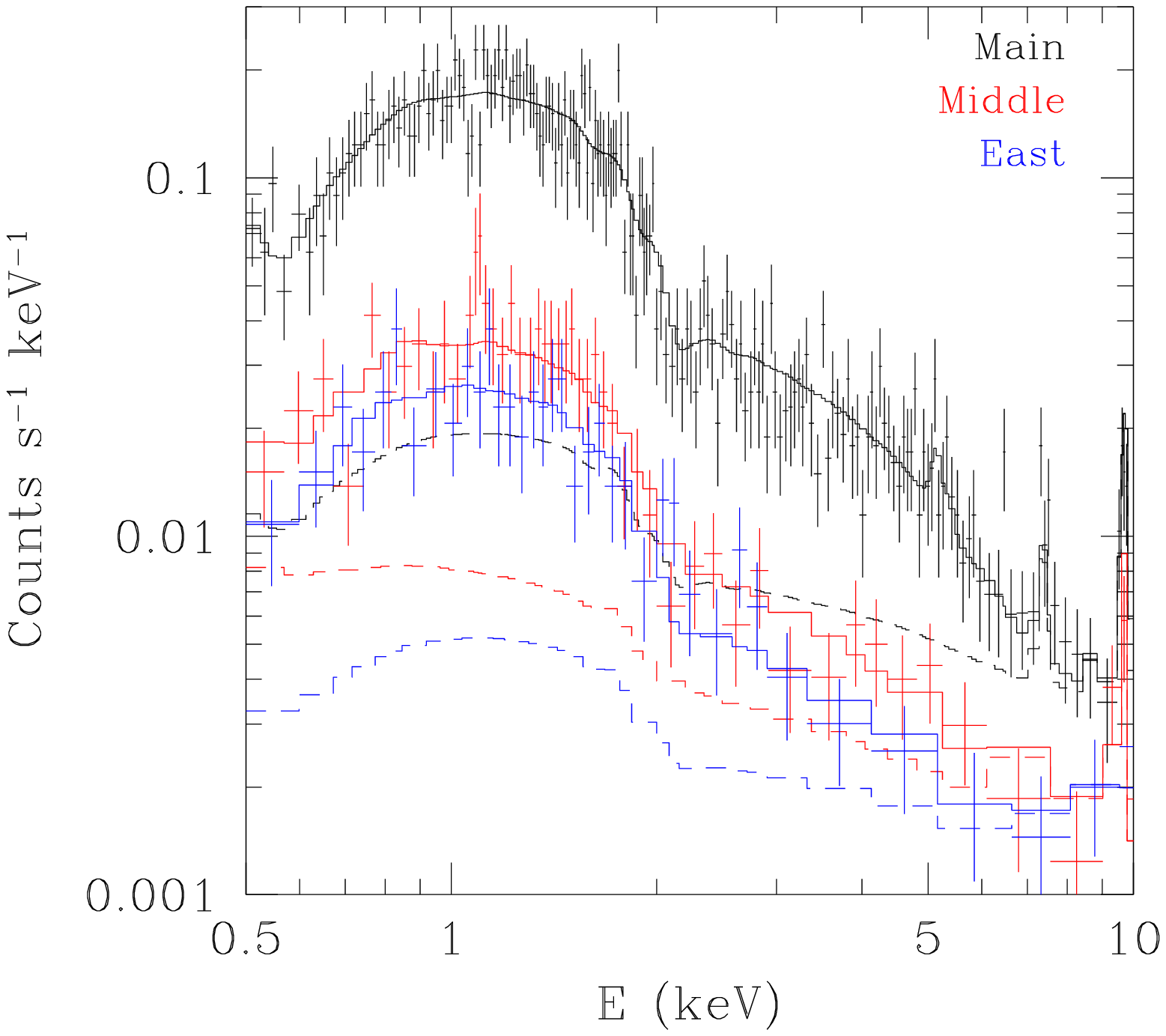}
\end{center}
\caption{From top to bottom, spectra for the Main, Middle, and East
clusters from \emph{Chandra}. Solid lines represent the best-fit
model, and dashed lines represent the contribution from the
background.} \label{fig:spectra_Chandra}
\end{figure}

The best-fit temperature values and 1-$\sigma$ statistical
uncertainties are given in Table \ref{tab:temps}. For the East and
Middle clusters, the metal abundance was held fixed to 0.3 times
solar. For the Main cluster, because of its higher statistical
level, this was allowed to vary yielding a best fit value of
$0.27\pm 0.15$ times solar (1-$\sigma$ error).  Redshifts were fixed
to the values mentioned above.  Figure \ref{fig:spectra_Chandra}
shows the best-fit spectra for these three clusters, where the
dashed line represents the contribution from the background.

\subsubsection{Analysis of XMM-Newton Data}

The initial data reduction steps for the \emph{XMM-Newton} data were
completed using the SAS software tools.  The data consists of
observations from the three EPIC instruments, MOS1, MOS2, and PN. To
model the non-X-ray background from charged particles and
instrumental fluorescence lines, we also obtained closed
observations (observations with the filter wheel in the closed
position) for each of the three EPIC instruments.  Thus we were able
to use an independent measure of the particle background, differing
from our \emph{Chandra} analysis.

Both our cluster data and the closed data were filtered by pattern
and energy range. We kept single and double pixel events and events
with pulse heights in the range of 300 to 12000 eV for MOS
observations. For PN observations, we only kept single pixel events
and events with pulse heights between 300 and 15000 eV.  We also
chose the most conservative screening criteria (excluding events
next to edges of CCDs and next to bad pixels, etc.). Our cluster and
closed EPIC data were also filtered for soft solar proton flares.

Figure \ref{fig:images} shows the 500-4500 ev band image of the EPIC
data after exposure correction and smoothing with a Gaussian of
$\sigma = 8^{\prime\prime}$.  Extraction regions for the clusters
and point sources were identified along with a background annulus
surrounding and away from any of the cluster regions (see Tables
\ref{tab:XMM_cluster_regions} and \ref{tab:XMM_points}). Spectra
were created of the cluster regions and background annulus for both
the closed and cluster data, as were spectral response and effective
area files. Resolved point sources were excluded from both the
source and background regions, and cluster emission was also
excluded from the background region.

Xspec was used to fit the spectra of the background annuli in the
closed and cluster data using a similar phenomenological model as
described for the \emph{Chandra} analysis. The closed data
background annulus was fit with several gaussians and three power
laws to model the non-X-ray background.  This best-fit model was
used as a starting model to fit the spectrum of the cluster data
background annulus, adding an absorbed power-law to model the
unresolved X-ray background and an absorbed soft thermal (mekal)
component to model the soft diffuse Galactic emission. The
absorption column density and photon index of the astrophysical
background model were fixed as mentioned above for the
\emph{Chandra} analysis, and the thermal component was given a fixed
plasma temperature of 0.2 keV and a solar metal abundance. We linked
the power-law norms of the particle background by the ratio of the
power-law norms in the closed background data. This kept the slope
and shape of the continuum fixed, but allowed the overall
normalization to vary.  The spectrum of the cluster data background
annulus was fit by allowing the normalizations of the astrophysical
background power-law and thermal component to vary as well as the
normalization of the non-X-ray particle background.

A best-fit joint model was created to fit the background annulus
spectra for the three instruments simultaneously. The parameters of
the astrophysical background model were kept in common between the
instruments, but the parameters of the particle background differed.
A second thermal component was added to the astrophysical background
model to better fit the spectra at energies below the aluminum
fluorescence line.  The normalizations, temperatures, and abundances
of the two thermal components were allowed to vary as well as the
unresolved X-ray background power law.

The source regions were fit drawing on the closed observations to
fit the non-X-ray background and the annulus spectra to fit the
astrophysical background.  The spectra of the cluster regions in the
closed observations were fit using the closed background annulus
best-fit model as a starting point.  The spectra of the source
regions in our data were fit using models starting where the
particle background normalization and the normalization of the
strongest fluorescence line of each cluster region were set equal to
those from the corresponding closed cluster region, scaled by the
ratio of the normalizations between the observed and closed
background annulus. The normalizations of the weaker fluorescence
lines were set equal to the normalization of the strongest line
scaled by the ratio of the weak-line norm to the strong-line norm in
the corresponding closed cluster region. The astrophysical
background model was taken from the best joint-fit model for the
background annulus, where the normalizations of the background
spectral components were scaled by the ratio of the exposure
integrated over the source and background regions. The source
spectra were modeled with a redshifted mekal thermal plasma model,
with the abundance equal to 0.3 times solar and the temperature and
normalization allowed to be free parameters.

We fit the spectra of the source regions using these starting models
by first fitting the normalization of the particle background using
only events greater than 10 keV, with the cluster model zeroed out.
With this normalization frozen, we fit the temperature and norm of
the cluster model using only events below 10 keV.  The particle
background norm was then refit using events larger than 10 keV and
this frozen cluster model.

Having fit for the particle background of each source region for
each instrument, the best-fit source models for the three
instruments were combined to allow for a joint fit. Again for the
joint fit, the astrophysical background model was kept the same for
each instrument except that the normalizations differed due to
different exposure scalings. This fit was done using the best-fit
particle background normalizations described above and those
differing by $\pm$ 1-$\sigma$ for each instrument. In this way, we
were able to model the systematic uncertainties arising from the
particle background subtraction.

\begin{figure}
\begin{center}
\vbox {\includegraphics[angle=-90,width=8cm]{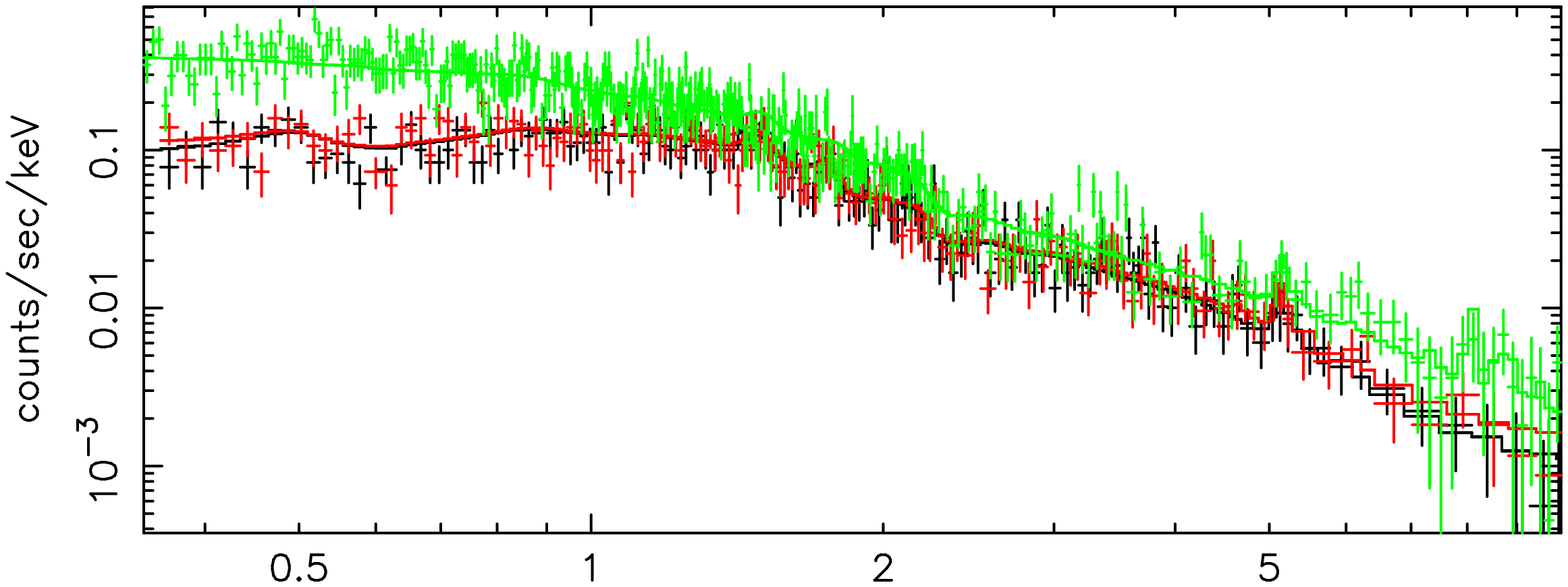} \vspace{0.5cm}
\includegraphics[angle=-90,width=8cm]{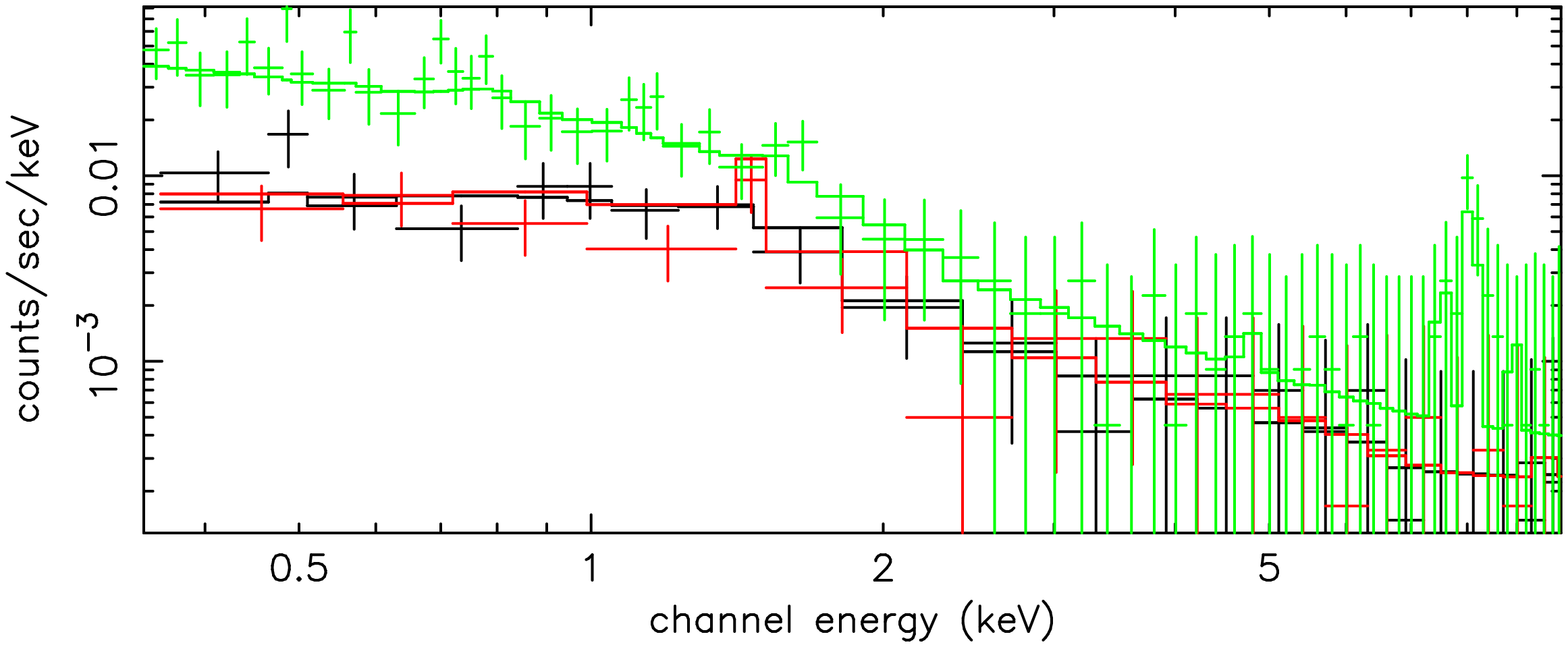}
} \vspace{-1.0cm}
\end{center}
\caption{Spectra for the Main cluster (top) and the East cluster
(bottom) from the three \emph{XMM-Newton} instruments. The Main
cluster has the highest, and the East cluster the lowest,
signal-to-noise of the four clusters in the pointing. Solid lines
represent the best-fit model, and black, red, and green colors
correspond to MOS1, MOS2, and PN instruments respectively.}
\label{fig:spectra1}
\end{figure}

\begin{figure}
\begin{center}
\vbox {\includegraphics[angle=-90,width=8cm]{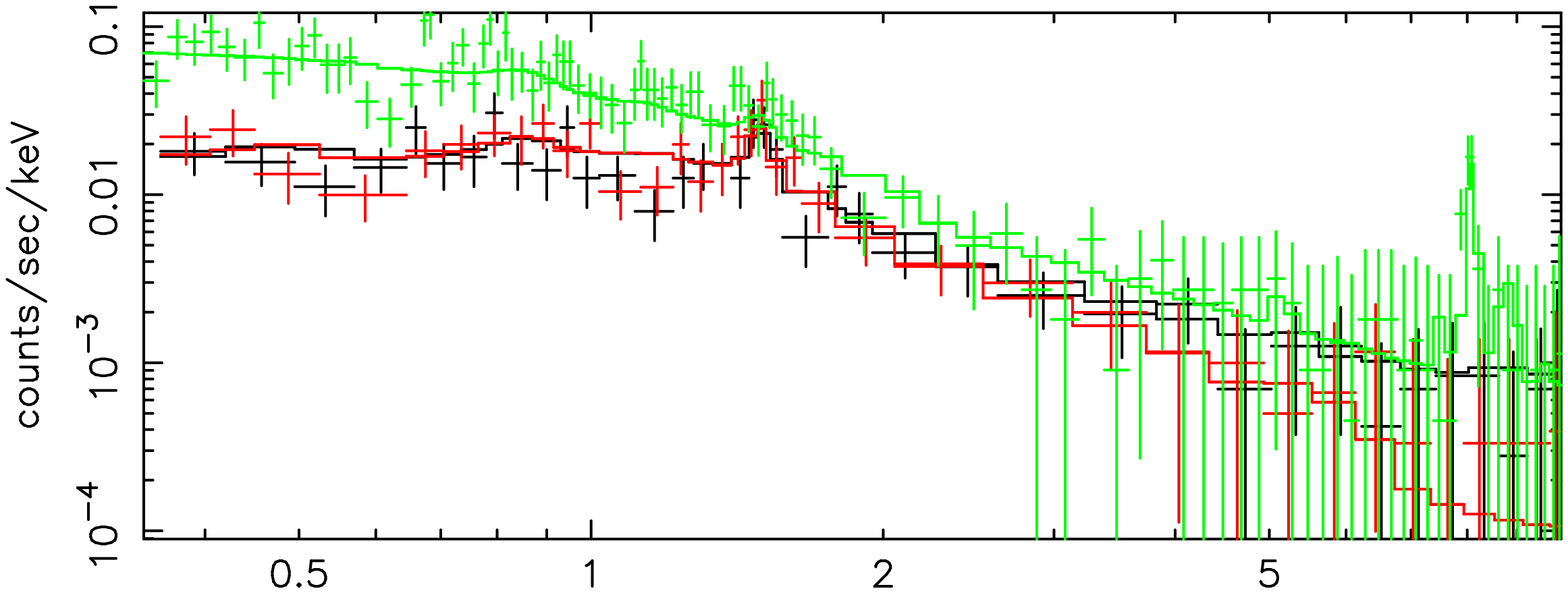} \vspace{0.5cm}
\includegraphics[angle=-90,width=8cm]{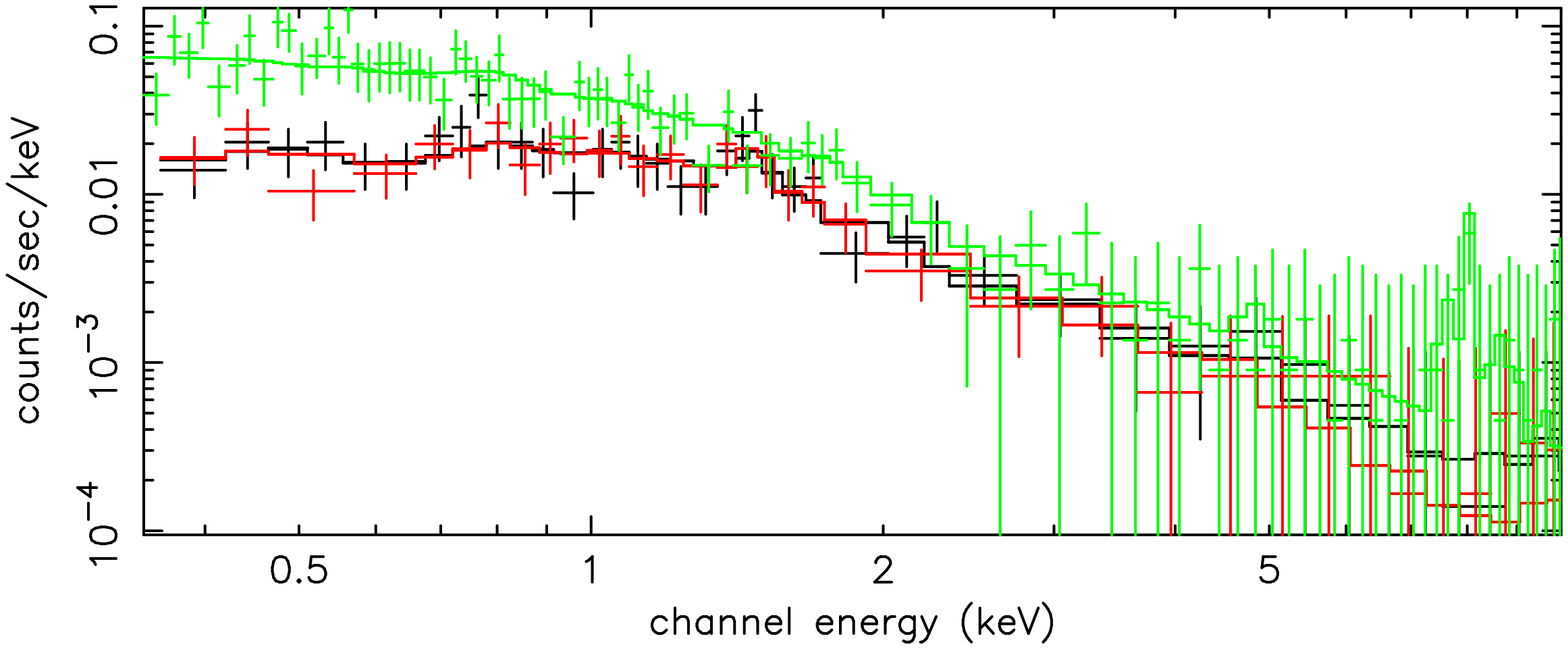}
} \vspace{-1.0cm}
\end{center}
\caption{Spectra for the Middle cluster (top) and the West cluster
(bottom) from the three \emph{XMM-Newton} instruments. Solid lines
represent the best-fit model, and black, red, and green colors
correspond to MOS1, MOS2, and PN instruments respectively.}
\label{fig:spectra2}
\end{figure}

The 1-$\sigma$ errors on the cluster temperature were obtained using
a delta fit statistic of 1.0 for the one interesting parameter.  The
resulting best-fit temperatures and error bars are given in Table
\ref{tab:temps}.  We find good agreement between the \emph{Chandra}
and \emph{XMM-Newton} best-fit temperatures given the statistical
and systematic uncertainties. The \emph{XMM-Newton} spectra for the
four main clusters and the best joint-fit models are shown in
Figures \ref{fig:spectra1} and \ref{fig:spectra2}.

\subsection{X-ray Surface Brightness Profiles}

A surface brightness profile was created by first determining the
surface brightness peak of each cluster.  This was done with
exposure corrected images in the 500-2000 eV band for each
instrument smoothed with a Gaussian of $\sigma = 8^{\prime\prime}$.
The vignetting function for the exposure maps was calculated at the
monochromatic X-ray energy of 1.25 keV for \emph{XMM-Newton} and 1
keV for \emph{Chandra}.

The unsmoothed images and exposure maps from the three EPIC
instruments and \emph{Chandra} were used to generate surface
brightness profiles.  Images and exposure maps from the separate
EPIC focal plane detectors were summed to create a joint \emph{XMM}
image and corresponding exposure map.  Exposure maps in all cases
were made including the effective area, for proper weighting of the
\emph{XMM} detectors and for ease of comparing the \emph{XMM} and
\emph{Chandra} profiles.

We chose 40 radial bins of 8$^{\prime\prime}$ each, centered around
the surface brightness peak of each cluster to find profiles that
extend out to 5.3'. In each radial bin, we summed the counts from
the \emph{Chandra}/joint \emph{XMM} image and divided this sum by
the total exposure in that bin from the corresponding exposure map.
We calculated the error on the surface brightness in each radial bin
by using the small count statistic
$(1+\sqrt{\rm{counts}+0.75})/\rm{exposure}$ \citep{Gehrels1986},
where units of exposure are given in sec cm$^2$ arcmin$^2$.

\begin{figure*}
\begin{center}
\vbox {\hbox {\includegraphics[angle=0,width=8cm]{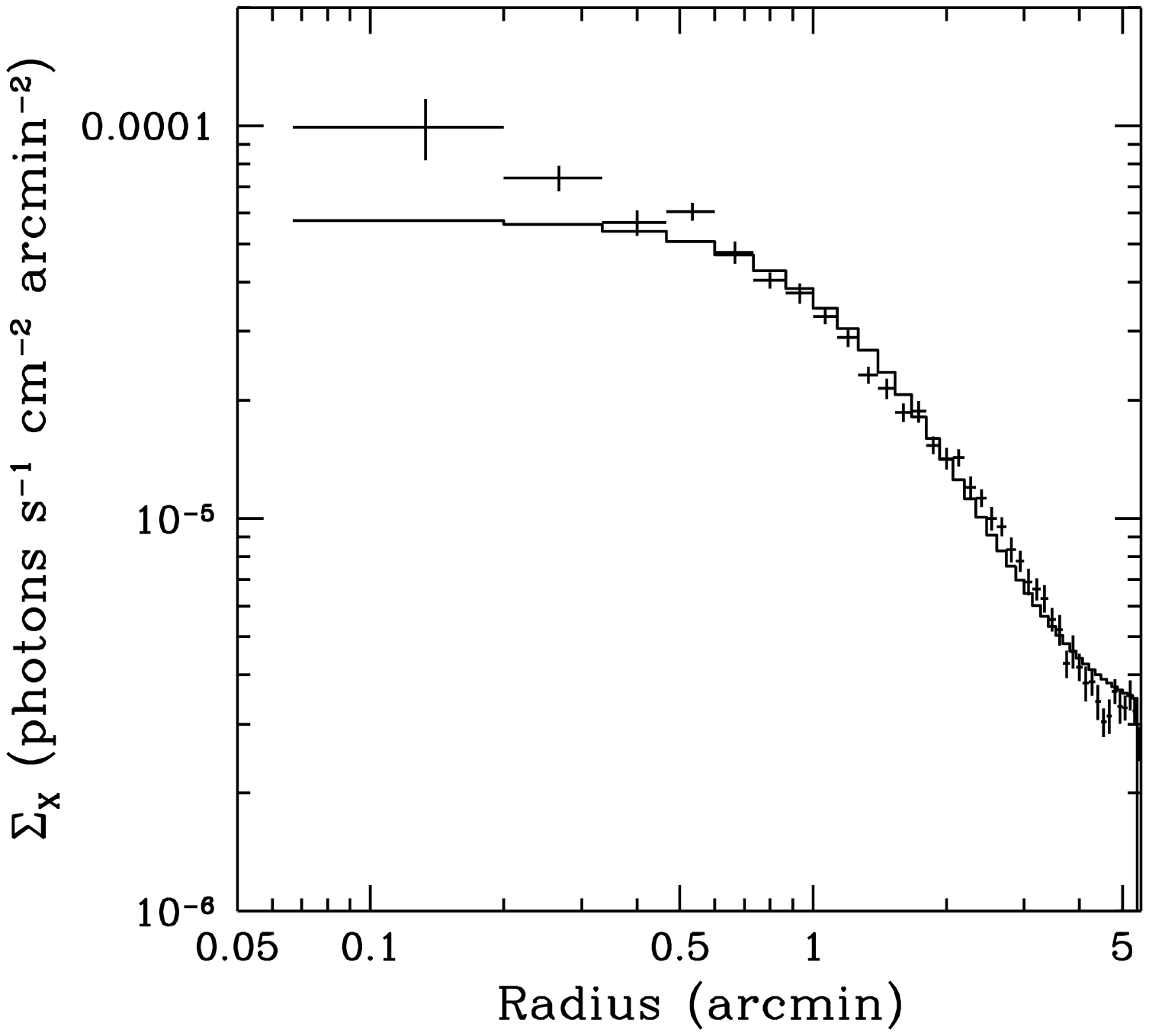}
\hspace{0.2cm}
\includegraphics[angle=0,width=8cm]{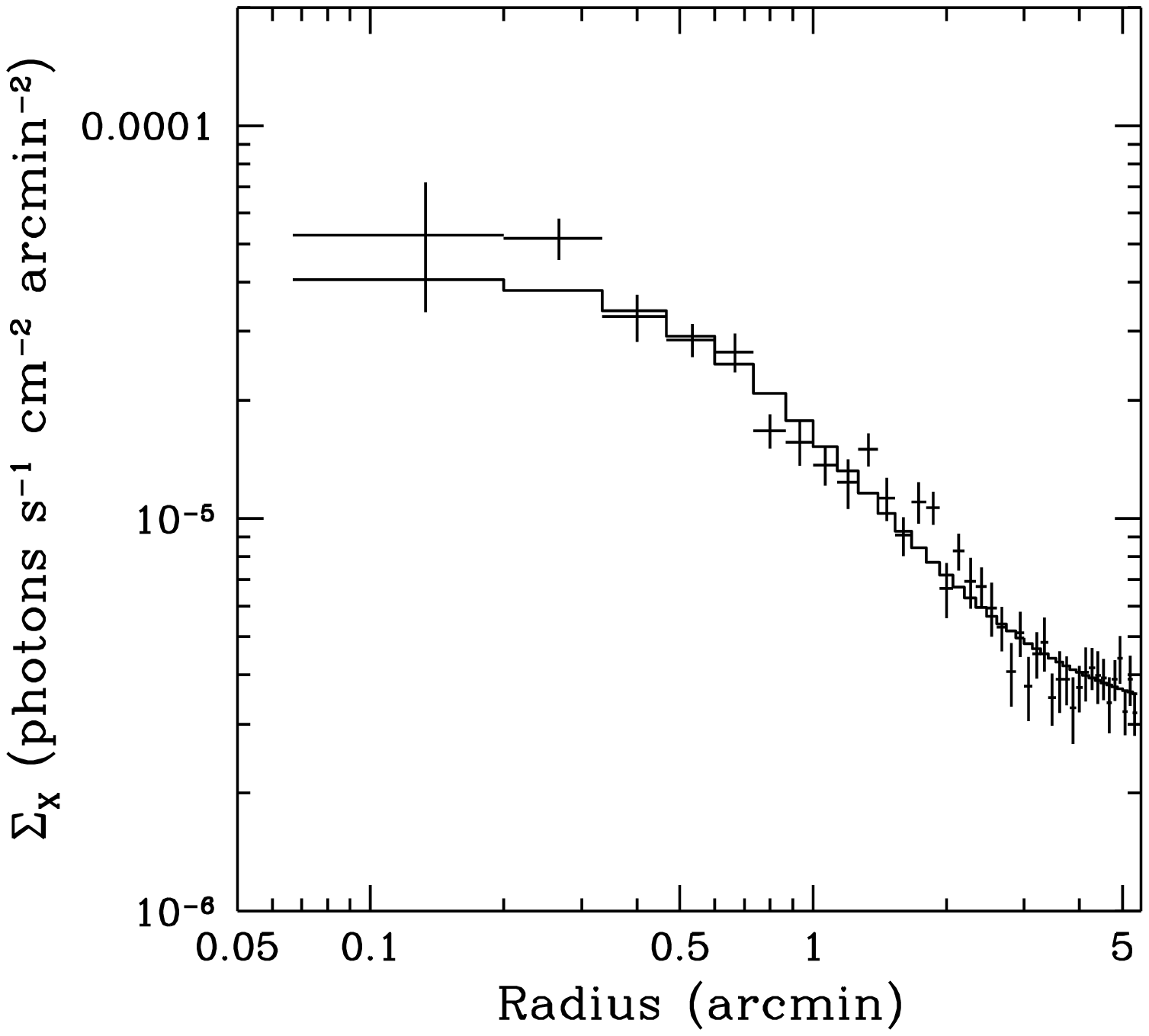}
} \hbox {\includegraphics[angle=0,width=8cm]{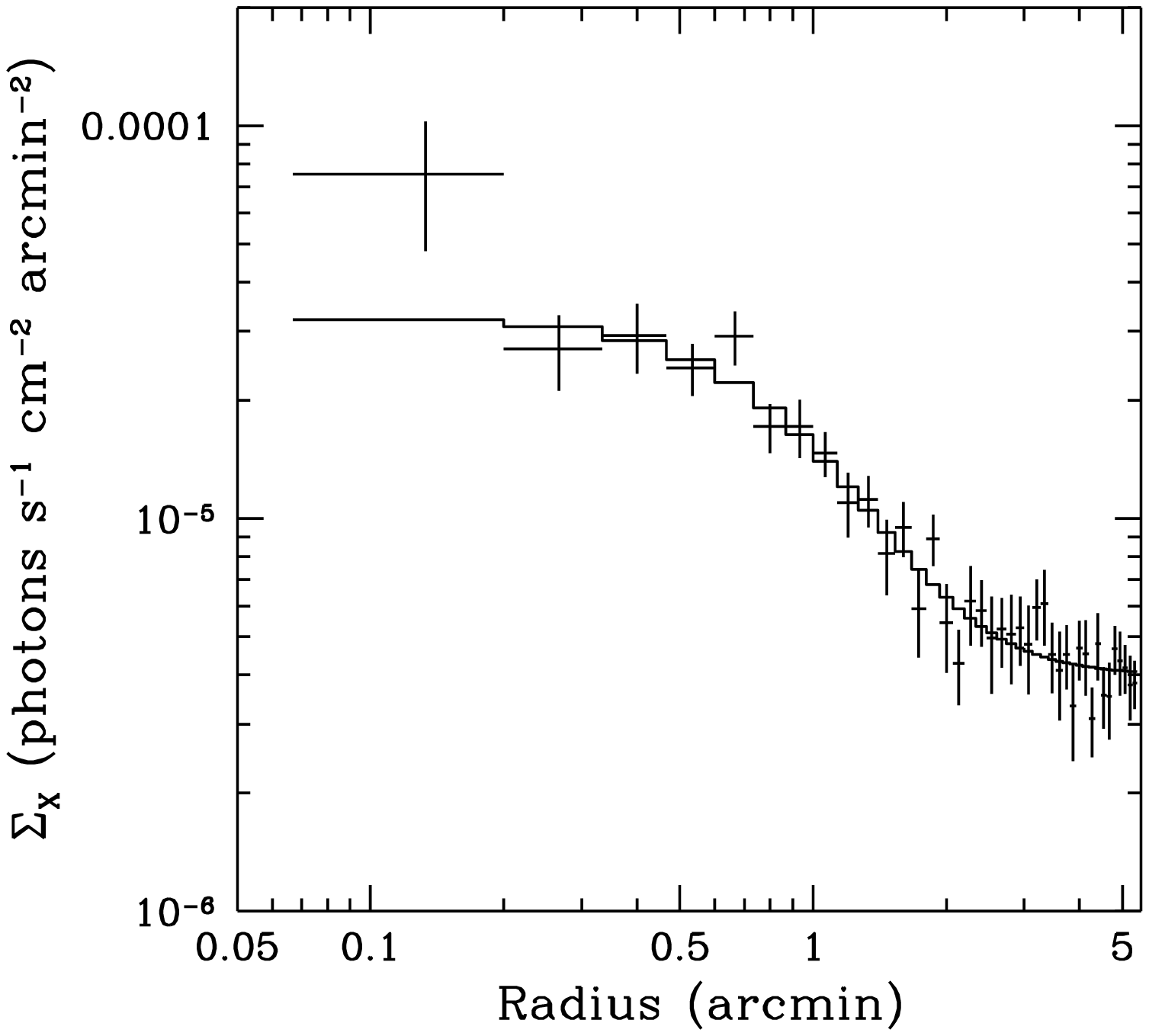} \hspace{0.2cm}
\includegraphics[angle=0,width=8cm]{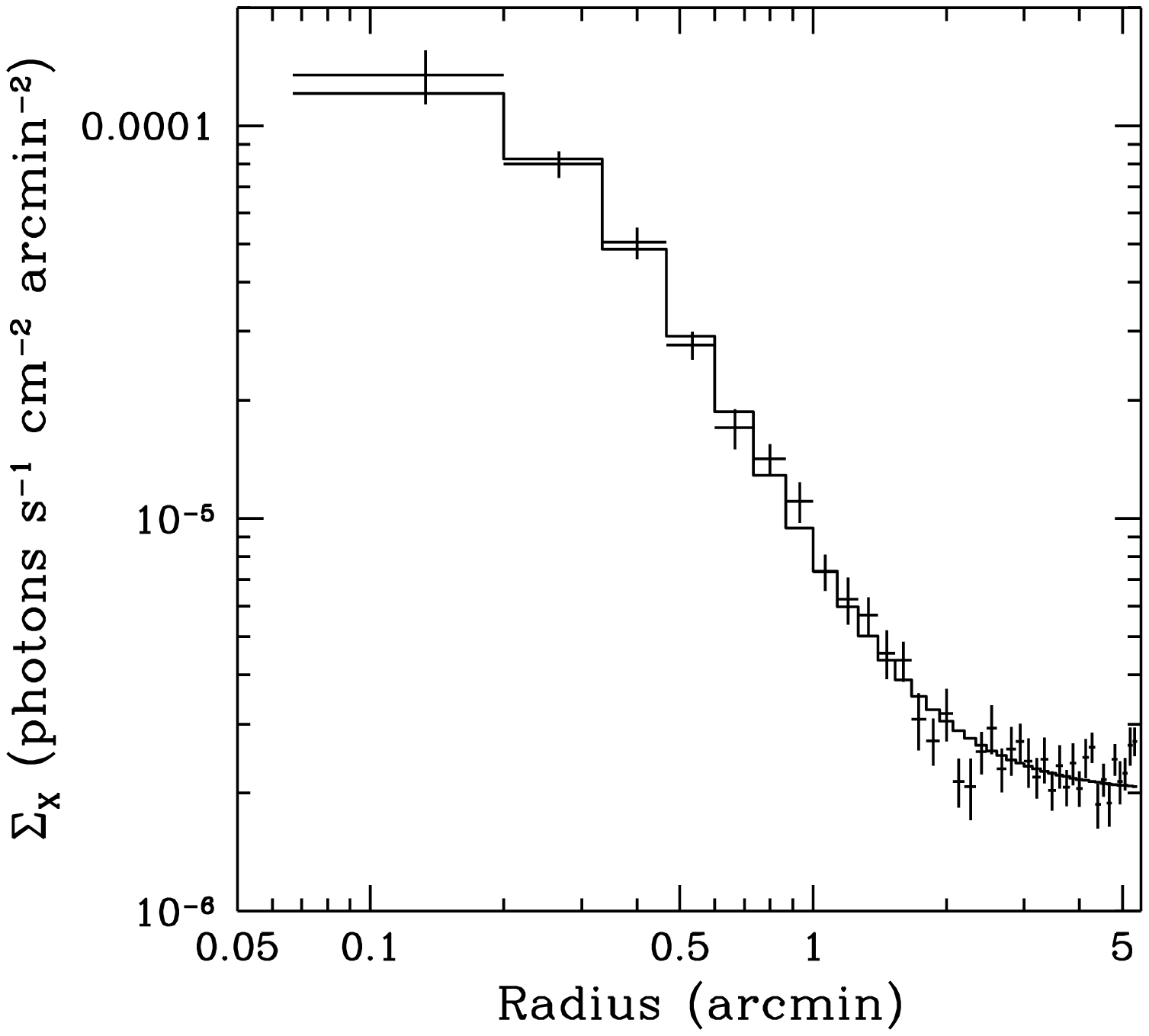}
}}
\end{center}
\caption{\emph{XMM-Newton} surface brightness profiles for the Main
(top, left), Middle (top, right), East (bottom, left), and West
(bottom, right) clusters and best-fit models. The energy band used
is 0.5-2 keV.} \label{fig:xmm_profiles}
\end{figure*}

\begin{figure*}
\begin{center}
\vbox {\hbox {\includegraphics[angle=0,width=8cm]{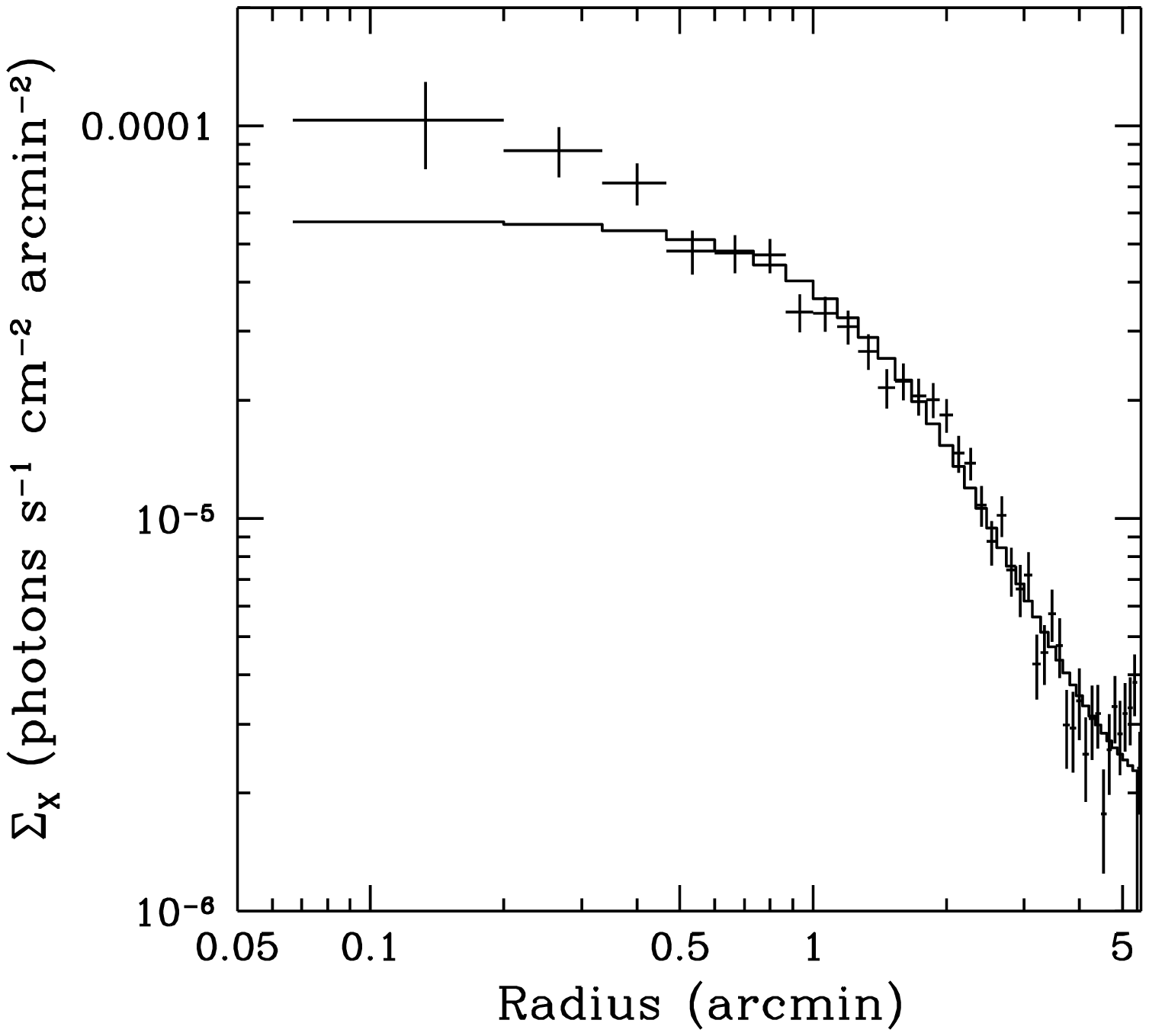}
\hspace{0.2cm}
\includegraphics[angle=0,width=8cm]{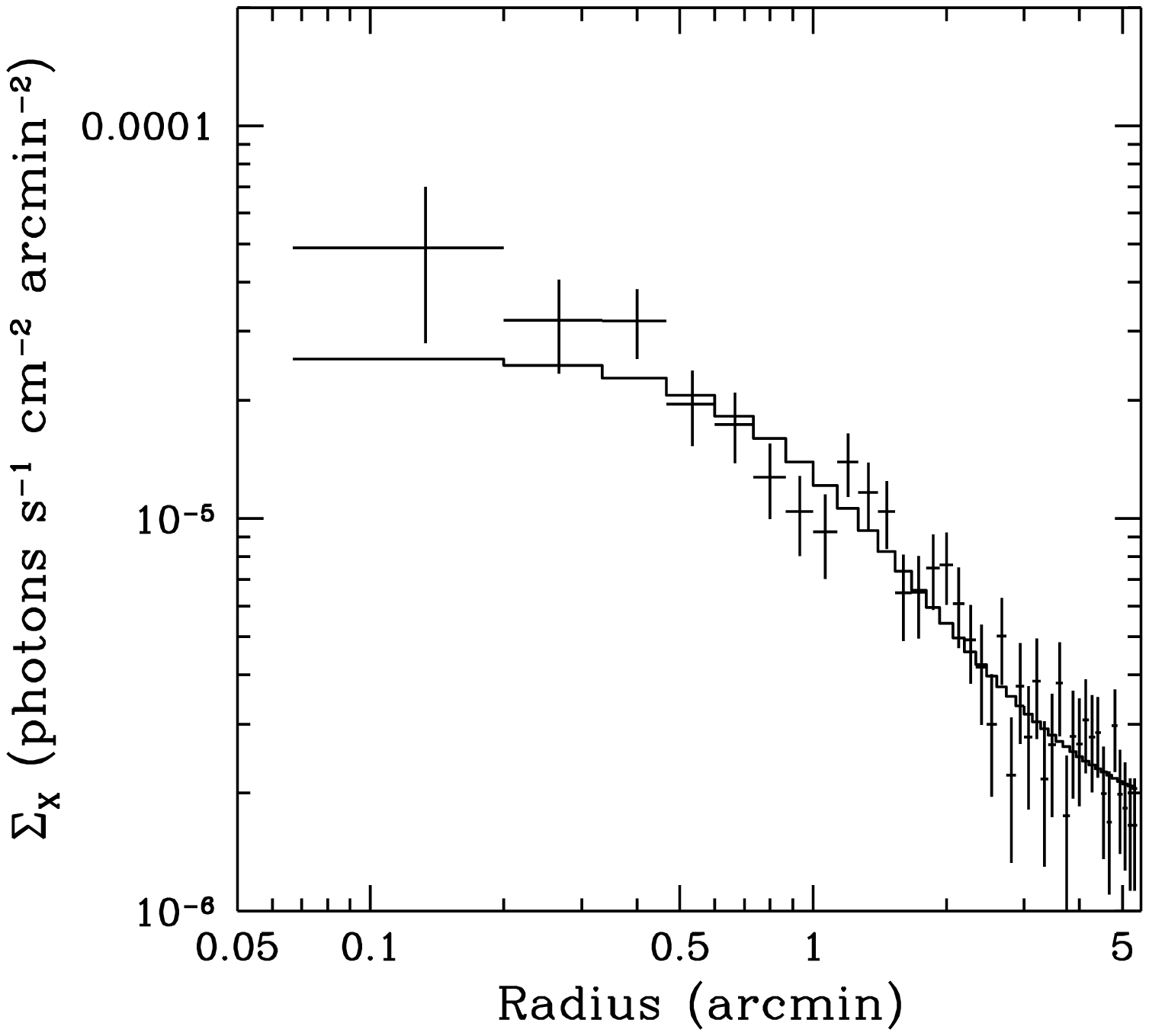}
}\hbox {\includegraphics[angle=0,width=8cm]{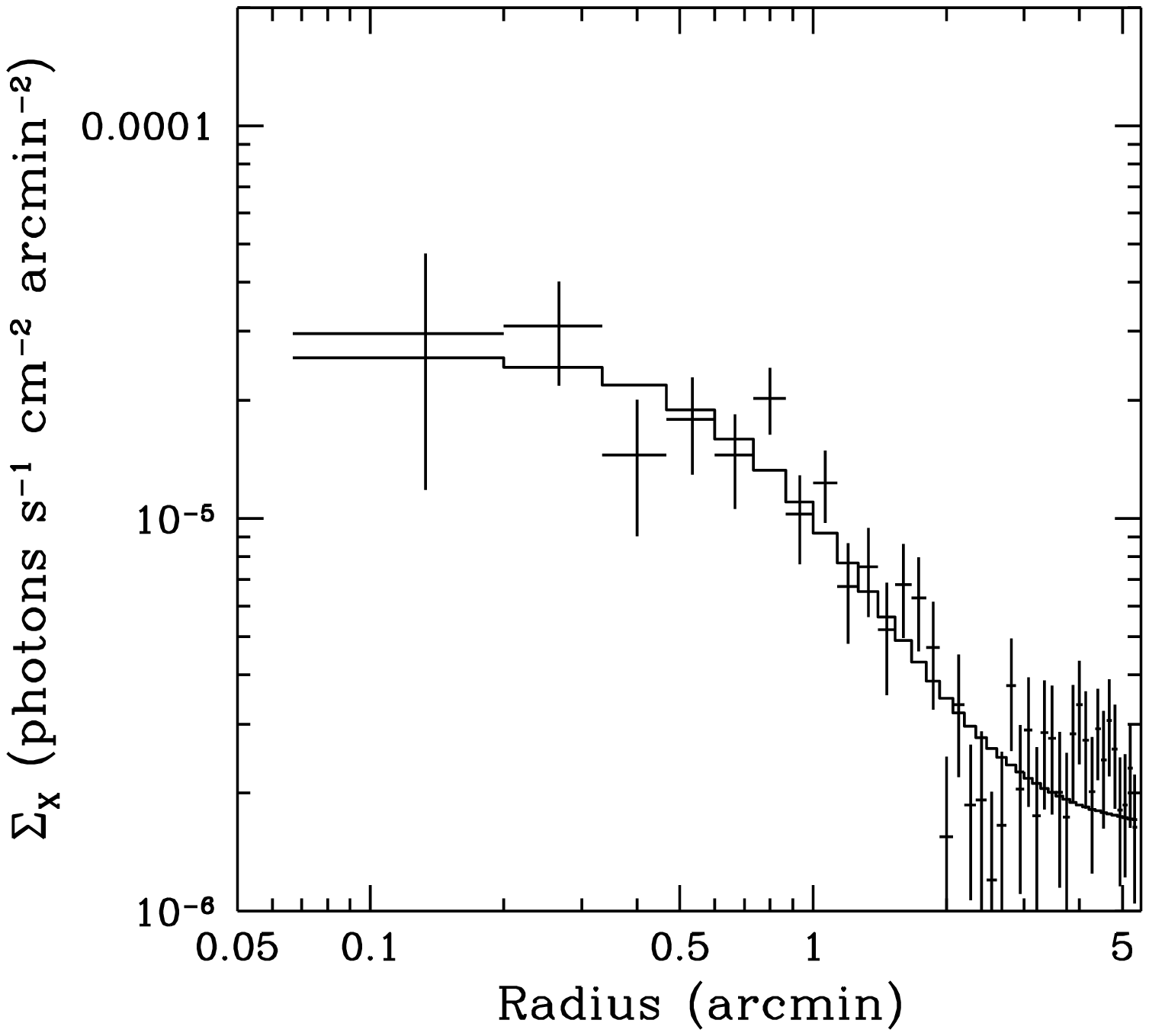} }}
\end{center}
\caption{\emph{Chandra} surface brightness profiles for the Main
(top, left), Middle (top, right), and East (bottom, left) clusters
and best-fit models. The energy band used is 0.5-2 keV.}
\label{fig:chandra_profiles}
\end{figure*}

Using the radial bins farthest from the surface brightness peaks, we
inferred the surface brightness due to the background.  We then fit
the surface brightness profiles to equation
\ref{eq:surface_brightness} by fixing the background and allowing
$\beta$, $r_c$, and the overall normalization to vary as free
parameters.  The profiles and best-fit models are displayed in
Figures \ref{fig:xmm_profiles} and \ref{fig:chandra_profiles}. Table
\ref{tab:beta_rcore} gives the $\beta$ and $r_c$ best-fit values for
each cluster along with their 1-$\sigma$ statistical error bars.
There is good agreement between the \emph{Chandra} and
\emph{XMM-Newton} best-fit $\beta$ and $r_c$ values given the
statistical uncertainties.

\begin{table*}
\begin{center}
\begin{tabular}{|c|c|c|c|c|}
\hline Cluster &\emph{XMM} counts &\emph{XMM} kT (keV)&\emph{Chandra} counts&\emph{Chandra} kT (keV) \\
\hline \hline
\hline East Cluster&505&$3.6^{+0.6+0.6}_{-0.5-0.7}$&300&$4.7^{+1.4}_{-1.0}$\\
\hline Middle Cluster&1135&$3.7^{+0.4+0.6}_{-0.3-0.4}$&380&$5.0^{+1.6}_{-1.1}$\\
\hline Main Cluster&8812&$6.3^{+0.3+0.4}_{-0.3-0.3}$&2400&$7.3^{+1.1}_{-0.7}$\\
\hline West Cluster&1163&$4.0^{+0.4+0.5}_{-0.3-0.5}$&0&...  \\
\hline
\end{tabular}
\end{center}
\caption{Integrated temperature estimates for the four clusters from
fits to the \emph{XMM-Newton} and \emph{Chandra} spectra. Note the
first \emph{XMM-Newton} error given is statistical and the second is
systematic due to background subtraction.} \label{tab:temps}
\end{table*}

\begin{table*}
\begin{center}
\begin{tabular}{|c|c|c|c|c|}
\hline Cluster &\emph{XMM} $\beta$ &\emph{XMM} $r_c$ (arcmin)&\emph{Chandra} $\beta$&\emph{Chandra} $r_c$ (arcmin) \\
\hline \hline
\hline East Cluster&$0.81^{+0.29}_{-0.15}$&$1.19^{+0.44}_{-0.24}$&$0.68^{+0.39}_{-0.13}$&$0.94^{+0.59}_{-0.29}$\\
\hline Middle Cluster&$0.51^{+0.05}_{-0.04}$&$0.71^{+0.20}_{-0.15}$&$0.56^{+0.13}_{-0.09}$&$0.99^{+0.49}_{-0.34}$\\
\hline Main Cluster&$0.87^{+0.06}_{-0.05}$&$1.82^{+0.15}_{-0.15}$&$0.88^{+0.12}_{-0.10}$&$2.01^{+0.34}_{-0.24}$\\
\hline West Cluster&$0.60^{+0.03}_{-0.02}$&$0.31^{+0.03}_{-0.03}$&...&...  \\
\hline
\end{tabular}
\end{center}
\caption{Best-fit $\beta$ and $r_c$ values for the four clusters
from fitting \emph{XMM-Newton} and \emph{Chandra} surface brightness
profiles assuming a $\beta$-model for the gas density.}
\label{tab:beta_rcore}
\end{table*}

\begin{table*}
\begin{center}
\begin{tabular}{|c|c|c|c|c|}
\hline Cluster &\emph{XMM} $r_{500}$&\emph{XMM} $M_{500}$&\emph{Chandra} $r_{500}$&\emph{Chandra} $M_{500}$ \\
&(Mpc)&($10^{14} M_{\odot}$)&(Mpc)&($10^{14} M_{\odot}$) \\
\hline \hline
\hline East Cluster&0.91&2.5&0.99&3.2\\
\hline Middle Cluster&0.84&1.7&1.01&3.0\\
\hline Main Cluster&1.38&7.6&1.48&9.5\\
\hline West Cluster&0.89&2.4&...&... \\
\hline
\end{tabular}
\end{center}
\caption{Best-fit \emph{Chandra} and \emph{XMM-Newton} mass
estimates within $r_{500}$ assuming an isothermal $\beta$-model.}
\label{tab:beta_model_masses}
\end{table*}

\section{Mass Estimates}

\subsection{X-ray Masses}

To determine each cluster's X-ray derived mass, we use equations
\ref{eq:hydrostatic}, \ref{eq:beta_model}, and \ref{eq:NFW_mass} and
our best-fit $\beta$ and $r_c$ values to predict the cluster
temperature profile for given values of the central density and
scale radius in the NFW profile.  The solution of the temperature
profile requires a boundary condition.  We assume the cluster
temperature to be 1.25 keV at 3.5 Mpc for each cluster but allow
this outer temperature and radius to vary generously over the range
of 0.5 to 2 keV and 2 to 5 Mpc and fold this uncertainty into our
error bars. We find that this variation in our boundary conditions
contributes a negligible amount (less than $4\%$) to our error
estimates.

To compare each cluster's X-ray and weak-lensing-derived masses in
an mutually consistent manner, we must assume the same matter
density profile for each method.  Since neither the X-ray nor the
weak-lensing data is deep enough to well constrain both $\rho_0$ and
$r_s$ in the NFW profile, we choose to estimate each cluster's
concentration parameter, $c\propto 1/r_s$, using results from
hydrodynamic simulations and an X-ray estimate of each cluster's
mass assuming isothermality. This mass estimate is accurate enough
to give a reasonable estimate of the concentration parameter since
the concentration is a slowly varying function of cluster mass.  The
concentration parameter is here defined as $r_{500}/r_s$, where
$r_{500}$ is the radius within which the cluster mean density is 500
times the critical density. We again use the isothermal case to
estimate $r_{500}$, which yields an estimate of $r_s$.  Since we are
primarily interested in the comparison between X-ray-derived and
weak-lensing-derived masses, the accuracy of the scale radius is
less important than the fact that we used the same scale radius in
deriving the masses using both methods. A change in the scale radius
may introduce a systematic shift in the masses but will not alter
significantly their ratio.

\subsubsection{Masses Assuming an Isothermal-$\beta$ Model}

We determine X-ray isothermal $\beta$-model mass estimates for these
clusters as an intermediate step to estimate the NFW scale radius
for each cluster.  If we assume that each cluster is isothermal with
a gas density given by equation \ref{eq:beta_model}, then equation
\ref{eq:hydrostatic} can be rewritten to give
\begin{eqnarray}
\lefteqn{M(r)=3\beta \frac{kTr}{G\mu
m_p}\Bigg[\frac{\big(r/r_c\big)^2}{1+\big(r/r_c\big)^2}\Bigg]}
\nonumber\\
& & {}= 1.13 \times 10^{15} \beta \frac{T_X}{10
\rm{\hspace{0.15cm}keV}} \frac{r}{\rm{Mpc}}
\Bigg[\frac{\big(r/r_c\big)^2}{1+\big(r/r_c\big)^2}\Bigg] M_\odot,
\end{eqnarray}
where we set $\mu = 0.59$ for the cluster gas \citep{Evrard1996}.
Using the best-fit $\beta$, $r_c$, and $T_X$ for each cluster from
\emph{Chandra} and \emph{XMM-Newton} observations, we calculate
$M_{500}$ for each cluster, which is the mass within $r_{500}$. The
best-fit $r_{500}$ and $M_{500}$ for each cluster from each X-ray
satellite are listed in Table \ref{tab:beta_model_masses}. This
table is without error bars since we only use these values to
estimate a reasonable $r_s$ for each cluster.

We use the above \emph{XMM} $r_{500}$ and $M_{500}$ values and
predictions for the concentration as a function of mass and redshift
derived from hydrodynamic cluster simulations \citep{Dolag2004} to
obtain an estimate of $r_s$.  The $c(M,z)$ relation used here is
from the \citet{Dolag2004} $\Lambda$CDM cosmological simulation, and
we converted our masses and radii to the definitions used in that
work to determine $r_s$.  This concentration relation has a
reasonable agreement with \emph{Chandra} observations of nearby
clusters \citep{Vikhlinin2006}.  We choose to use \emph{XMM}
$\beta$-model masses since we have \emph{XMM} observations for all
four clusters.  Table \ref{tab:central_densities} lists the best
estimates of $r_s$. We hold this scale radius fixed and keep it in
common when deriving masses from weak-lensing and X-ray methods and
focus on the relative difference in mass estimates using these two
methods.

\subsubsection{Masses Assuming an NFW Profile}

Given an estimate of $r_s$, $\beta$, and $r_c$ for each cluster, we
vary $\rho_0$ and calculate the emission-weighted, projected,
average temperature within an aperture (as discussed in \S
\ref{sec:profiles}). The apertures for each cluster and instrument
are given in Tables \ref{tab:Chandra_cluster_regions} and
\ref{tab:XMM_cluster_regions}. We match the predicted temperatures
to our measured temperatures in Table \ref{tab:temps} to find the
best-fit central densities.  The 1-$\sigma$ errors on $\rho_0$ are
calculated by varying the measured $T_X$, $\beta$, and $r_c$
parameters within their 1-$\sigma$ error bars and treating $\beta$
and $r_c$ as correlated variables.  The variation in the boundary
condition introduces a negligible error on $\rho_0$ as mentioned
above.  Best-fit $\rho_0$ measurements and 1-$\sigma$ errors for
\emph{Chandra} and \emph{XMM} observations are given in Table
\ref{tab:central_densities}.  There is good agreement between the
two instruments, and we calculate a weighted average to give a
combined X-ray $\rho_0$ for each cluster, which we also list in
Table \ref{tab:central_densities}.  Given $\rho_0$ and $r_s$, X-ray
masses are calculated using equation \ref{eq:NFW_mass} and listed in
Table \ref{tab:masses}.  These masses are calculated using the
combined X-ray $\rho_0$ for all the clusters except the West
cluster, for which only the \emph{XMM} $\rho_0$ is available.

\subsection{Weak-Lensing Masses}

The weak lensing mass estimates are based on the full DLS exposure
time of 18 ks in $R$ and 12 ks in $BVz'$, rather than the partially
complete imaging used for cluster discovery in \citet{Wittman2006}.
Otherwise, image processing was as described in detail in that
paper. The $R$ data were taken in good seeing conditions (FWHM
$<0.9''$) and are used to measure the galaxy shapes.  The A781
complex spans two contiguous DLS ``subfields'' or pointing centers.
The FWHM of the final $R$ images, after circularizing the PSF and
co-adding 20 exposures, is $0.78''$ and $0.74''$ for the two
subfields involved.  The $BVz'$ data are used only to provide color
information for photometric redshifts.

We extracted shear catalogs using a partial implementation of
\citet{Bernstein2002}.  This implementation appeared as the ``VM''
method in \citet{Heymans2006}, which compared different weak-lensing
methods on a set of simulated sheared images.  After correcting for
stellar contamination which was present in that dataset but not
here, the VM method yielded 89\% of the true shear in those images.
In this work, we divide the VM results by 0.89 to more closely
approximate the true shear.  However, we recognize that this
correction factor is likely to be data-dependent, and we therefore
assign a systematic error of 10\% to the shear calibration.

We derived photometric redshifts using BPZ \citep{Benitez2000} with
the HDF prior.  We optimized the templates using a subset of the
SHELS \citep{Geller2005} spectroscopic sample and the procedure of
\citet{Ilbert2006}.  Complete details are discussed elsewhere
(Margoniner et al., in preparation).  To assess the accuracy of
these photometric redshifts, we turn to an independent spectroscopic
sample, consisting of all redshifts in DLS field F2 in the
literature, as tabulated by the NASA/IPAC Extragalactic Database. We
find 328 galaxies with spectroscopic redshifts that match our
cleaned photometry (unsaturated, not near bright stars, etc),
spanning the redshift range 0.02---0.70.  The resulting rms
photometric redshift error per galaxy is $0.047(1+z)$, with a bias
of $-0.017(1+z)$, and no catastrophic outliers.

Because shear is nonlocal and mass maps tend to be highly smoothed,
the presence of one clump may affect the apparent mass density of
another clump.  This is true of the \citet{Wittman2006} maps, and we
eliminate that effect here by simultaneously fitting axisymmetric
NFW profiles to the four X-ray positions.  The model fitting takes
into account the full three-dimensional position (RA, DEC, z) of
each source galaxy. The per-galaxy imprecision in $z$ is not
important because the lensing kernel is very broad, and because each
galaxy is a very noisy estimator of the shear: a 0.1 shift in source
redshift changes the modeled shear by much less than the per-galaxy
shear error.  For each NFW model, RA and DEC were fixed by the X-ray
position, $z$ was fixed by the spectroscopy, and $r_s$ was fixed to
the value used for the X-ray fitting.  Thus only one parameter,
$\rho_0$, was fit for each model.  \citet{Wright2000} give
expressions for shear induced by an NFW profile.

Of the $\sim$350,000 galaxies in the 2$^\circ\times$2$^\circ$ DLS
field containing A781, we limited the fit to galaxies within
15\arcmin\ of a clump center, for computational efficiency and
because more distant galaxies may be influenced more by other
clusters than by those in the A781 complex.  We also cut on
photometric redshift, because cluster members scattering to higher
redshift would reduce the estimated shear (slightly, because of
their low inferred distance ratio), while cluster members scattering
to lower redshift would have no effect.  We did fits with cuts at
$z_{\rm phot} > 0.35$ (just behind the richer, lower-redshift
clumps) and $z_{\rm phot} > 0.62$ ($3\sigma$ beyond the
higher-redshift clumps), and found a difference of $\leq 0.2\sigma$
in the fitted parameters.  The lenient (strict) cut yielded 30137
(22173) galaxies, or 23 (17) arcmin$^{-2}$ over 1320 arcmin$^2$,
although $\sim$10\% of this area was masked due to bright stars.  We
adopt the strict cut to avoid any question of contamination.  The
resulting source catalogs show no increased density near the clump
centers.

Because shear from an NFW profile is linear in $\rho_0$, we used
singular value decomposition (SVD) as described in
\citet{Press1992}). This solves the general linear least squares
problem in one pass, with no iteration required.  The galaxies were
given equal weights in the fit, because the VM shear method does not
assign weights to galaxies. However, the importance of a galaxy in
determining the fit still depends on its position and redshift,
through the model's dependence on position and redshift.  We then
corrected for the fact that the observable in weak lensing is not
the shear $\gamma$, but the reduced shear ${\gamma/(1-\kappa)}$
(where $\kappa$ is the convergence) as follows. We computed the
convergence of the best-fit model at the location of each source
galaxy, constructed a reduced-shear model, redid the linear fit, and
iterated.  In the first iteration, this resulted in a $\sim$5\%
correction to the fit parameters.  In the second iteration, the
correction was only $\sim$0.2\%, much smaller than the fitted
parameter statistical uncertainties, and therefore the reduced-shear
fit was deemed to have converged.

The fitted $\rho_0$'s and their uncertainties are listed in Table
\ref{tab:central_densities}.  The uncertainties output by the SVD
routine were confirmed by 1000 bootstrap resampling realizations.
Although the fit is not $\chi^2$-driven, we can define a $\Delta
\chi^2$ statistic to see how much of the variance in the data is
explained by the model.  There are 44326 degrees of freedom because
each galaxy has two shear components.  For each component of shear,
we compute the rms without any fit and define that as the
uncertainty associated with each galaxy.  This results in an initial
$\chi^2$ of 44326, which is decreased by 40.4 when the
four-parameter model is subtracted off.  The chance of this
happening randomly is $<10^{-7}$ for Gaussian distributions.  The
remaining variance is due mostly to the intrinsically random
distribution of galaxy shapes (shape noise) and shape measurement
errors, although a small amount may be attributed to additional
structure in the field as indicated by the following mass maps.

\begin{figure}
\begin{center}
\includegraphics[angle=0,width=8cm]{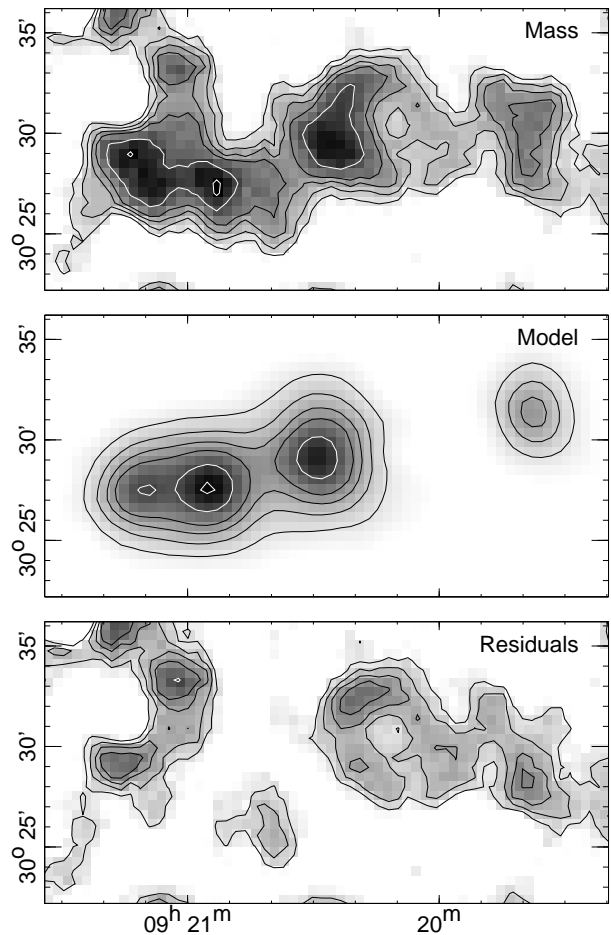}
\end{center}
\caption{Lensing data, model, and residuals.  Top: mass map of the
area.  Grayscale and contours are in arbitrary units, but the same
units are used for all the panels.  Middle: best-fit lensing mass
model. Bottom: mass map made from residual shear after subtracting
off the quadruple NFW profile fit shown in the middle panel.
\label{fig-massmaps}}
\end{figure}

We show the fit visually in Figure~\ref{fig-massmaps}.  The top
panel shows a mass map made using the method described in
\citet{Wittman2006} (originally from \citet{Fischer1997}), the
middle panel shows a similar map made from the model shears, and the
bottom panel shows a map made from the residual shear after
subtracting off the fit.  The Main, Middle, and East clumps have
been mostly subtracted, but the West clump has not been well
modeled.  There is also an unmodeled mass clump just northwest of
the East clump (RA $\approx$ 9:21, Dec $\approx$ 30:33), which
appears to have some associated galaxies. We caution that the mass
map is a sanity check rather than a quantitative indicator of
goodness-of-fit, because it does not fold in source redshift
information.

\begin{table*}
\begin{center}
\begin{tabular}{|c|c|c|c|c|c|c|}
\hline Cluster &Estimated $r_s$ &\emph{XMM} $\rho_0$&\emph{Chandra} $\rho_0$&Combined X-ray $\rho_0$&DLS $\rho_0$ \\
&(Mpc)&($10^{-26}$ g/cm$^3$)&($10^{-26}$ g/cm$^3$)&($10^{-26}$ g/cm$^3$)&($10^{-26}$ g/cm$^3$) \\
\hline \hline
\hline East&0.37&$3.9^{+1.0}_{-1.0}$&$5.2^{+1.6}_{-1.2}$&$4.4\pm 0.8 $&$4.4\pm 1.3\pm 0.6$\\
\hline Middle&0.31&$5.2^{+1.1}_{-0.7}$&$7.3^{+2.3}_{-1.7}$&$5.8\pm 0.9$&$6.8\pm 1.5\pm 0.8$\\
\hline Main&0.60&$3.1^{+0.3}_{-0.2}$&$3.5^{+0.5}_{-0.4}$&$3.2\pm 0.2$&$2.2\pm 0.4\pm 0.3$\\
\hline West&0.33&$6.4^{+1.0}_{-0.9}$&...&$6.4\pm 1.0$&$4.0\pm 1.7\pm 0.4$ \\
\hline
\end{tabular}
\end{center}
\caption{Estimated scale radii of NFW mass profiles and best-fit NFW
central densities from \emph{Chandra}, \emph{XMM-Newton}, and DLS.}
\label{tab:central_densities}
\end{table*}

\begin{table*}
\begin{center}
\begin{tabular}{|c|c|c|c|c|}
\hline Cluster &X-ray $r_{500}$&X-ray $M_{500}$&Weak-lensing $r_{500}$&Weak-lensing $M_{500}$ \\
&(Mpc)&($10^{14} M_{\odot}$)&(Mpc)&($10^{14} M_{\odot}$) \\
\hline \hline
\hline East Cluster&$0.74^{+0.06}_{-0.07}$&$1.8^{+0.5}_{-0.5}$&$0.73^{+0.11}_{-0.13}$&$1.8^{+0.9}_{-0.8}$\\
\hline Middle Cluster&$0.76^{+0.06}_{-0.06}$&$1.7^{+0.4}_{-0.4}$&$0.82^{+0.10}_{-0.12}$&$2.0^{+1.0}_{-0.7}$\\
\hline Main Cluster&$1.09^{+0.04}_{-0.04}$&$5.2^{+0.3}_{-0.7}$&$0.89^{+0.10}_{-0.12}$&$2.7^{+1.0}_{-0.9}$\\
\hline West Cluster&$0.79^{+0.06}_{-0.06}$&$2.2^{+0.5}_{-0.4}$&$0.60^{+0.15}_{-0.14}$&$1.1^{+0.8}_{-0.7}$ \\
\hline
\end{tabular}
\end{center}
\caption{X-ray and shear masses within $r_{500}$ for the four
clusters assuming an NFW matter density profile.} \label{tab:masses}
\end{table*}

In Table \ref{tab:central_densities}, weak-lensing errors are of two
types, the first is statistical and the second is systematic.  For
the statistical errors, we performed bootstrap resampling to
estimate the covariance of the cluster mass estimates.  The masses
of neighboring clusters (in projection) are anticorrelated, because
the observed shear in a region must include the sum of the model
shears from the neighbors. The errors given here for each cluster
are after marginalizing over the allowed values for the other
clusters. Therefore, the error on the total mass of all the clusters
would be smaller than the quadrature sum of the errors given here.
Also, the covariance will affect the comparison of the lensing
$\rho_0$ for one of the four clusters to that for any of the others.

Systematic errors include shear calibration, source redshift
calibration, mass sheet degeneracy, and residual cluster member
contamination.  We assign a shear calibration systematic uncertainty
of 10\% as explained above.  We explored the effects of source
redshift errors by fitting to different source redshift ranges.  We
find that only if we include photometric redshifts beyond 1.4, where
the 4000 \AA\ break redshifts out of the \emph{z}$^\prime$ filter,
do the results change by as much as 10\%. Even in that case, the
results appear to change randomly rather than systematically, with
some clumps increasing in fitted mass and others decreasing.  Given
these results, and the small bias in photometric redshifts when
compared to the spectroscopic sample from the literature, a
systematic error of 5\% in mass due to redshift calibration errors
is very conservative.

We estimate residual cluster member contamination by examining the
source galaxy areal density around the richest (Main) clump.  There
is an excess in the central 2\arcmin\ radius. Comparing the redshift
distribution in that area to a control annulus, we find 58 excess
galaxies at redshifts 0.35--0.42, presumably cluster members with
$>1\sigma$ photometric redshift errors.  In the fit, all galaxies
were weighted equally, so their relative importance is determined by
their proximity (in three dimensions) to where the model reduced
shear is large. These galaxies are near the projected center of the
lens, but at low inferred distance ratio.  We compute their
effective weight as the square of the model reduced shear, and
compare their total weight to the rest of the galaxies within
5\arcmin\ of the Main clump.  The summed weight of the interloping
galaxies is only 0.010 times that of the valid source galaxies,
which presumably resulted in a 1\% underestimate of the mass, much
smaller than the other systematics.

The last systematic involves mass sheet degeneracy and projection of
unrelated structures near the line of sight. {\it If} our assumption
of an NFW profile is correct, then mass sheet degeneracy is not an
issue, because we are fitting the profile rather than empirically
determining departures from a baseline.  Measuring shear to as large
a radius as possible would help check the profile assumption, just
as it would help reduce mass sheet degeneracy.  However, going to
large radii increases the chance of including some unrelated
structure projected near the line of sight. \citet{Metzler2001}
(hereafter M01) characterized this by measuring lensing masses of
clusters in large-scale numerical simulations as a function of data
set radius.  The truncation radius used in the fitting here,
30$^\prime$, corresponds to a transverse separation of about 5 Mpc,
for which M01 find a scatter of about 7\%.  (Note that M01's
systematic offset in the population is not relevant here because
profile fitting is less biased than their aperture densitometry.)
Scatter in the population becomes a systematic when considering a
single cluster.  However, by simultaneously fitting multiple clumps,
this systematic is probably already greatly reduced.  We empirically
test this effect by varying the truncation radius and find that the
results can change by up to 5\%.  We therefore assign a systematic
of 5\% due to this effect. The larger systematic is likely to be in
the profile assumption. This systematic is likely to be on the same
order as the mass sheet degeneracy systematic incurred if the
profile assumption were dropped, which is quoted as $20\%$ by
\citet{Bradac2004}.

In summary, the systematics include $20\%$ for the profile
assumption and/or mass sheet degeneracy, $10\%$ for shear
calibration, $5\%$ for projections, and $5\%$ for photometric
redshift errors.  We include only the latter three systematic errors
in Table \ref{tab:central_densities}, where we assume an NFW profile
accurately describes the matter density.  This allows us to compare
the X-ray and weak-lensing $\rho_0$ values.  In table
\ref{tab:masses}, we present a summary of the $r_{500}$ and
$M_{500}$ values, where the the weak-lensing statistical and
systematic error bars on $\rho_0$ are added in quadrature. It should
be kept in mind that the absolute masses of the clusters depend on
the profile assumption.

\section{DISCUSSION}

Comparison of the X-ray and weak-lensing $\rho_0$ values indicates
that these values are in agreement for the East, Middle, and West
clusters. This agreement suggests that line-of-sight mass
contributions have not significantly biased the weak-lensing
measurements.

For the Main cluster, the X-ray-derived $\rho_0$ is higher than that
from weak-lensing by about 2$\sigma$.  The X-ray images of the Main
cluster suggest it may be undergoing a merger with a subcluster (see
Figure \ref{fig:images}), and thus may be out of hydrostatic
equilibrium.  From the literature it is not obvious whether cluster
mergers are expected to bias X-ray mass estimates high or low.
Weak-lensing observations of 22 X-ray bright clusters at
$0.05<z<0.31$ found X-ray temperatures higher by $40-75\%$ than
those inferred from weak-lensing, for clusters with $T_X > 8$ keV
\citep{Cypriano2004}.  The largest discrepancy in this sample was
for the two clusters with the highest X-ray temperatures ($T_X \sim
13$ keV), which both show signs of being out of dynamical
equilibrium.  It is a reasonable extrapolation to presume that all
the clusters with $T_X > 8$ keV in their sample are unrelaxed, with
the temperature of their intracluster medium boosted by shocks due
to in-falling groups and mergers with other clusters.

However, recent hydrodynamic simulations indicate that unrelaxed
cluster temperatures should be lower than those for relaxed clusters
of the same mass \citep{Kravtsov2006,Nagai2007}.  The reasoning for
this is that over the course of a merger, the mass of the system
increases faster than the conversion of the kinetic energy of the
merging systems into thermal energy of the intracluster medium
\citep{Kravtsov2006}. It has also been suggested from hydrodynamic
simulations of merging clusters that X-ray mass estimates based on
hydrostatic equilibrium can be biased high close to core-crossing,
where a temperature boost occurs due to shocks, but can be biased
low just before and after this temperature spike
\citep{Poole2006,Puchwein2007}.  This latter scenario is possibly
supported by recent X-ray and weak-lensing observations of 10 X-ray
luminous clusters at $z\sim 0.2$ \citep{Zhang2007, Bardeau2007}. The
X-ray observations of this sample were carried out with
\emph{XMM-Newton}, and the weak-lensing is from ground-based imaging
with the CFH12k camera on CFHT.  These authors report that four out
of these ten clusters have consistent X-ray and weak-lensing mass
estimates.  Of the six clusters that show a mass discrepancy, half
have higher and half have lower X-ray as compared to weak-lensing
masses.

It is possible that the Main cluster in the A781 cluster complex is
close to core-crossing as our X-ray mass estimate is biased high. It
is also possible that a selection effect is occurring in the sample
of \citet{Cypriano2004} where the clusters with $T_X > 8$ keV are in
a state of high temperature boosting and thus close to
core-crossing. Deeper X-ray observations of this A781 cluster and
further comparisons of weak-lensing and X-ray mass estimates of
known merging clusters will help to clarify the biases expected from
dynamically unrelaxed systems.

The West cluster did not appear in our original shear maps, and we
confirm here that it is a low significance weak-lensing detection
(1-2$\sigma$).  It was detected in the X-ray by chance as it fell
within the same \emph{XMM-Newton} pointing as the other three
clusters. However, weak-lensing and X-ray data yield consistent mass
estimates.  We measure a very small core radius for this cluster
($r_c = 0.31'$), consistent with most of the emission coming from a
compact core.  Northwest of the East cluster, we also find an
enhancement in both the galaxy distribution and lensing signal, and
we find some indication of X-ray emission from that region. Further
X-ray observations would allow us to study these lower mass systems
in more detail.

Based on the limited information available, the velocity dispersion
values appear broadly consistent with the East, Middle, and Main
clusters having nearly equal masses \citep{Geller2005}. This agrees
with the similarity in weak-lensing masses we present here (see
Table \ref{tab:masses}).  A detailed comparison of the cluster
masses derived above with the velocity distributions is beyond the
scope of this work. Study of the velocity distribution of the
potentially merging component (Main cluster) could shed light on the
epoch and geometry of the merger (see \citet{Gomez2000}).

\section{CONCLUSION}

Many cluster surveys will take place over the next few years at
several different wavebands.  Already considerable samples of X-ray
and optically selected clusters have been compiled.  In addition,
sizeable microwave and shear-selected samples are close at hand.
These surveys can probe with precision the growth of structure over
cosmic time, and thereby open a new window on cosmology.  However,
the two main hurdles to overcome are relating cluster observables to
mass and characterizing the sample selection. By comparing
weak-lensing and X-ray observations and mass estimates for clusters
in the DLS shear-selected sample we hope to understand the
systematic biases in both mass estimation methods and modes of
cluster selection.

An analysis of the top shear-ranked mass distribution in the DLS
sample reveals a complex of four clusters, the largest of which can
be identified as A781.  The four clusters are at distinctly
different redshifts, as determined by optical spectroscopy, and the
X-ray images suggest three are dynamically relaxed while the largest
cluster appears to be merging with a small subcluster.  Masses from
both X-ray and weak-lensing observations were determined assuming an
NFW profile for the matter density.  Since neither sets of
observations were deep enough to well constrain both the central
density and scale radius of each cluster NFW profile, we estimated
the scale radii using X-ray-derived isothermal $\beta$-model mass
estimates and a relation describing concentration as a function of
mass and redshift derived from cosmological hydrodynamic
simulations.  For each cluster profile, the same scale radius was
used to determine the best-fit X-ray and weak-lensing central
densities.  We focus on the difference in central densities derived
with each method as the central density scales linearly with mass.

We find that three out of the four clusters show agreement between
their X-ray and weak-lensing derived central densities. The fourth
and largest cluster has an X-ray derived central density higher than
that derived from weak-lensing by about 2$\sigma$.  This discrepancy
is most likely due to the cluster's disrupted dynamical state.
Recent weak-lensing observations of X-ray selected clusters and
hydrodynamic simulations leave some ambiguity about whether
dynamical disruption via cluster mergers biases X-ray mass estimates
high or low.  The direction of the bias may be related to the stage
of the merger, e.g., whether it is close to core-crossing or more
advanced. Deeper X-ray observations of this cluster to better
resolve the merger and further comparisons between weak-lensing and
X-ray-derived masses of known merging clusters will shed greater
light on this issue.

Initial steps are being made by many groups to overcome the above
mentioned hurdles regarding the use of clusters as precision tracers
of structure growth.  Our collaboration, for example, has
\emph{Chandra} and \emph{XMM-Newton} data on a larger sample of DLS
shear-selected clusters that we will be reporting on in future
publications.  Hopefully all these efforts will open a new window
through which we can understand our universe.

\acknowledgments The DLS has received generous support from Lucent
Technologies and from NSF grants AST 04-41072 and AST 01-34753.
Space-based follow-up of shear-selected clusters in the DLS is
supported by NASA grant number NNG05GD32G to UC Davis and
\emph{Chandra} grant GO3-4173A, \emph{XMM-Newton} grant NAG5-13560,
and NASA LTSA grant NAG5-11714 to Rutgers. Observations were
obtained at Kitt Peak National Observatory and the W. M. Keck
Observatory. This work also made use of the Image Reduction and
Analysis Facililty (IRAF), the NASA/IPAC Extragalactic Database
(NED), and the NASA Astrophysics Data System (ADS).  JPH would like
to acknowledge the late Leon Van Speybroeck (who was the PI of the
Chandra observation analyzed here) for his selfless dedication to
the development of Chandra (then known as AXAF), for his wisdom and
patience as a scientific mentor, and for the exceptional humanity he
displayed throughout his life.

{\it Facilities:} \facility{CXO}, \facility{XMM},
\facility{Mayall(Mosaic), \facility{Keck:I(LRIS)}}

\bibliographystyle{apj}


\end{document}